\documentclass[aps,prl,twocolumn,superscriptaddress,10pt,longbibliography,floatfix,noeprint]{revtex4-2}
\usepackage{amsmath,amssymb,amsfonts,mathtools,amsthm}
\usepackage{graphicx}
\usepackage[dvipsnames]{xcolor}
\usepackage{times}
\definecolor{myrefcolor}{rgb}{0.067,0.5,0.5}
\usepackage[colorlinks=true,linkcolor=myrefcolor,citecolor=myrefcolor,
            urlcolor=myrefcolor,breaklinks]{hyperref}
\DeclareMathOperator{\Tr}{Tr}
\newcommand{\ket}[1]{\lvert#1\rangle}
\newcommand{\bra}[1]{\langle#1\rvert}
\newtheorem{theorem}{Theorem}
\newtheorem{proposition}{Proposition}

\newcommand{\figincl}[2][\columnwidth]{\includegraphics[width=#1]{#2}}
\graphicspath{{./}}
\begin{document}
\title{Quantum Mpemba effect from logical-information exposure in monitored circuits}

\author{Bochen Huang}
\affiliation{Institute of Physics, Beijing National Laboratory for Condensed Matter Physics, Chinese Academy of Sciences, Beijing 100190, China}
\affiliation{School of Physical Sciences, University of Chinese Academy of Sciences, Beijing 100049, China}

\author{D. L. Zhou}
\email{zhoudl72@iphy.ac.cn}
\affiliation{Institute of Physics, Beijing National Laboratory for Condensed Matter Physics, Chinese Academy of Sciences, Beijing 100190, China}
\affiliation{School of Physical Sciences, University of Chinese Academy of Sciences, Beijing 100049, China}

\author{Chenfeng Cao}
\email{caocf@hku.hk}
\affiliation{HK Institute of Quantum Science and Technology, The University of Hong Kong, Hong Kong, China}

\begin{abstract}
In the Mpemba effect, a state initially farther from stationarity overtakes
one that starts closer. After one cycle in a monitored Clifford chain, an
unencoded Bell-pair state has lower mean entanglement entropy than a
scrambled reference prepared from fewer pairs. The gap grows linearly with
chain length. This reversal is driven by logical-information exposure:
a stabilizer state loses one ebit per independent logical parity resolved by
the record. Scrambling within each half can spread this information over many
sites while preserving every Schmidt coefficient. We prove that with independent exact unitary two-design encoders, any
pure reference ensemble with no greater mean initial entropy can lead in mean
entropy by at most two ebits at any time. If the stationary entropy is subextensive, the unencoded reversal
is a quantum Mpemba effect at large sizes. We further certify this effect
at 128 and 256 qubits using stationary-entropy bounds.
\end{abstract}

\maketitle
\raggedbottom

\emph{Introduction.}---Can encoding alone suppress a reversal of entanglement order
under local monitoring? Consider Bell pairs joining mirror-image sites
across a chain's center, with product qubits next to the cut. We prepare
a reference from fewer Bell pairs by scrambling within each half. After
one cycle of random Clifford gates and local measurements, the mean
entanglement entropy of the unencoded state falls below that of the
reference by an amount proportional to chain length. The unencoded state
never regains an extensive lead in mean entropy. By contrast, uniformly scrambling
the initially more entangled state within each half preserves every
Schmidt coefficient. We prove that the reference's mean entropy cannot exceed that of the encoded state by more than two ebits at any time under the same monitored evolution.

The quantum Mpemba effect (QME) concerns a preparation initially farther
from stationarity overtaking a closer one
\cite{Mpemba1969Cool,Lu2017Nonequilibrium,Ares2025The}. For the pair
above, a bound on the stationary entropy lets us test this distance
ordering directly. Quantum examples include Markovian relaxation \cite{Carollo2021Exponentially,Nishanth2026DisorderBlind},
symmetry restoration
\cite{Ares2023Entanglement,Joshi2024Observing,Turkeshi2025Quantum,Liu2024Symmetry,Yamashika2026Symmetric},
and monitored dynamics, conditioned on no jumps
\cite{DiGiulio2025Measurement} or resolved along trajectories
\cite{Ganguly2026Measurement}. The effect also occurs without global
symmetries \cite{Bhore2025Without} and has been studied through complexity,
resources and entropy differences
\cite{Aditya2026Complexity,Summer2026Resource,Ares2025Simpler}.
Entanglement itself exhibits anomalous dissipative relaxation
\cite{Liu2026SuddenDeath,Bao2026Entanglement,Benjadi2026Exponential}.
In a dissipative Dicke model, Das \emph{et al.}\ reported different
entanglement decay for locally rotated pure states with identical Schmidt spectra
\cite{Das2025Role}. In a monitored circuit, the record lets us trace this dependence
to the information acquired by the detectors. After scrambling, the bit
revealed by measuring one end of a bare Bell pair may be accessible
only by reading several sites together.

In monitored circuits, information extraction competes with scrambling
\cite{Skinner2019Measurement,Li2018Quantum,Li2019Measurement}.
Their error-correcting structure protects logical information
\cite{Gullans2020Dynamical,Choi2020Quantum,Fan2021Self,Li2021Statistical},
and monitored-code criteria identify the measured logical operators
that destroy this protection \cite{LeeYoshida2026Randomly,Watanabe2026Homological}.
Initial-state dependence near the monitored transition has also been
studied for volume-law and product preparations
\cite{Wang2026Relaxation}.
Here we identify a first-readout threshold set by the redundancy of a
scrambled preparation and distinct from the circuit's measurement-induced
transition. A scrambled state can retain nearly all of
its entanglement at this step even when the stationary state obeys an
area law.

\begin{figure}[!t]
\centering\figincl{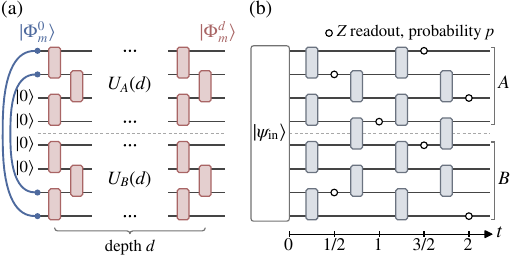}
\caption{Fixed-spectrum encoding and common monitoring.
(a) Intra-half Clifford circuits of depth $d$ transform the
$m$-Bell-pair state $\ket{\Phi_m^0}$ into $\ket{\Phi_m^d}$ while
preserving its Schmidt spectrum. The dashed line marks the bipartition.
(b) The input $\ket{\psi_{\rm in}}$ evolves through gate layers, each
followed by independent probability-$p$ $Z$ readout (open circles).
An odd-bond and an even-bond layer form one cycle.}
\label{fig:setup}
\end{figure}

\smallskip
\emph{Model and preparations.}---An open chain of $L=2n$ qubits, with
even $n$, undergoes an odd-bond layer on $(1,2),(3,4),\ldots$, followed by
an even-bond layer on $(2,3),(4,5),\ldots$ [Fig.~\ref{fig:setup}(b)].
All gates are independent uniform two-qubit Cliffords. Each layer is
followed by independent probability-$p$ projective $Z$ readout without
feedback. The first readout is at $t=1/2$ and each even-bond readout completes a cycle. We average the pure-trajectory half-chain entropy
$S_A=-\Tr\rho_A\log_2\rho_A$ over preparations, gates, measurement
locations and Born outcomes.

The unencoded partial rainbow $\ket{\Phi_m^0}$ contains Bell pairs
$(a,L+1-a)$ for $a\le m$. The remaining qubits next to the cut are in $\ket{0}$.
Independent intra-half encoders produce
$\ket{\Phi_m^d}=(U_A(d)\otimes U_B(d))\ket{\Phi_m^0}$, where one depth
unit is a simultaneous nearest-neighbor matching in each half [Fig.~\ref{fig:setup}(a)].
Each encoded half is maximally mixed on a code with $m$ logical qubits.
The $n-m$ checks inherited from the product qubits give the encoding
its redundancy. We write $\ket{\Phi_m^{\rm orb}}$ and
$\ket{\Phi_m^{\rm des}}$ for draws obtained from $\ket{\Phi_m^0}$ using
independent uniform half-chain Cliffords (yielding the uniform local Clifford orbit)
\cite{Fattal2004Entanglement,Smith2006Typical,Dahlsten2006Entanglement}
and independent exact unitary two-design encoders, respectively.

The main comparison fixes $m=3L/8$ and uses $\ket{\Phi_{m'}^{\rm orb}}$
with $m'=21L/64$ as the reference. Its local Clifford invariance permits
exact counting. Write $\overline S_d(t)$ for the depth-$d$ mean entropy
and $\overline S_{\rm ref}(t)$ for the reference mean. Then
$D_d(t)=\overline S_{\rm ref}(t)-\overline S_d(t)$ is positive when the
initial entropy order has reversed. For orbit and two-design preparations,
we use subscripts $\mathrm{orb}$ and $\mathrm{des}$ in place of $d$.
Reversing the gap $m-m'=3L/64$ requires an extensive difference in
the entropy changes. At $p=1/5$, this gap lies
between the unencoded first-readout loss and the much smaller loss after scrambling.

\begin{figure}[!t]
\centering
\figincl{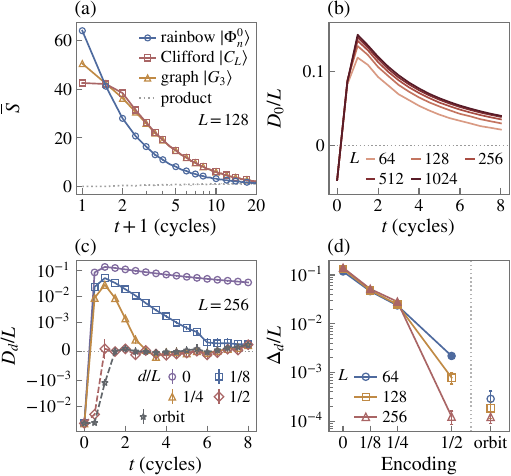}
\caption{Reversal of entanglement ordering and its control by encoding at
$p=0.20$. (a) Mean half-chain entropy $\overline S$ at $L=128$ for the complete rainbow
$\ket{\Phi_n^0}$, depth-$L$ Clifford $\ket{C_L}$, three-regular graph
$\ket{G_3}$, and product states. The logarithmic axis shows $t+1$.
(b--d) Fixed-spectrum comparisons use $m=3L/8$ and the common reference
$\ket{\Phi_{m'}^{\rm orb}}$ with $m'=21L/64$.
(b) Lines show $D_0(t)/L$ from 1000 paired trajectories per size.
Panels (c,d) use a separate sample.
(c) $D_d(t)/L$ at $L=256$ for depths $d/L=0,1/8,1/4,1/2$ and the orbit,
with a symmetric logarithmic vertical axis.
(d) Peak difference $\Delta_d/L$, where $\Delta_d=\max_tD_d(t)$ on
$t=0,0.5,\ldots,256$. Error bars show standard errors in (a,d) and
pointwise 95\% confidence intervals in (c).
Filled (open) markers in (d) indicate resolved (unresolved) crossings
(SM, Sec.~\ref{sec:SM_methods}).}
\label{fig:phenomenon-mechanism}
\end{figure}

\smallskip
\emph{Reversal and its control.}---Figure~\ref{fig:phenomenon-mechanism}(a)
illustrates the phenomenon at $L=128$ with the complete rainbow
$\ket{\Phi_n^0}$ \cite{Vitagliano2010Rainbow,Ramirez2014Rainbow},
states $\ket{C_L}$ prepared by a depth-$L$ random Clifford circuit,
graph states $\ket{G_3}$ on three-regular graphs, and product states. The rainbow starts with
the largest entropy and falls below the Clifford and graph states
early in the evolution.

In panel (a) the initial entanglement and its encoding both vary.
Panels (b--d) fix both the Schmidt spectrum of the initially more entangled
state and the reference $\ket{\Phi_{m'}^{\rm orb}}$ at each $L$.
All traces start at $D_d(0)/L=-3/64$.
In panel (b), $D_0(t)/L$ peaks at one cycle for each size, rising
from $0.118$ at $L=64$ to $0.149$ at $L=1024$.

Increasing only the encoding depth of the initially more entangled state
suppresses the reversal in panels (c,d). At $L=256$, the estimated peak
falls from $36.0$ ebits
without encoding to about $0.03$ ebit for both $d=L/2$ and the uniform
orbit. At this size, neither of these two preparations has a resolved
crossing within 256 cycles. The Supplemental Material (SM) \cite{SupplementalMaterial}
gives the statistical criteria and results for all sizes
(Sec.~\ref{sec:SM_methods}).

\smallskip
\emph{What a measurement record reveals.}---The unencoded state already
falls below the reference in mean entropy at the first readout. The two
halves have yet to interact, so any loss across the cut comes from the
readout itself. Consider first a set
$J\subseteq A$. Its Born-averaged entanglement loss $\kappa_J$ is the
information its outcome record $O_J$ carries about the opposite half,
\begin{equation}
 \kappa_J=I(O_J:B).
 \label{eq:record-information-main}
\end{equation}
Conditioning on the record leaves the joint state pure. The entropy
lost across the cut is therefore exactly the reduction in the mean
entropy of $B$ when the outcomes on $A$ are known
\cite{Groenewold1971Problem,Ozawa1986Information,Holevo1973Bounds,Buscemi2008Global}.
For stabilizer states, the recorded Pauli products that commute
with all local checks, modulo those checks, are the logical parities
resolved by the measurement \cite{LeeYoshida2026Randomly}.
Each independent parity distinguishes orthogonal sectors on $B$ and
removes one ebit. We call this \emph{logical-information exposure}.
For bilateral readout, we first count the information gained on $A$,
then count the additional information gained on $B$ in the state
conditioned on the first record.
The rank formula and the general information identity are given in
End Matter~\ref{em:record}.

Measuring a Pauli operator that anticommutes with a local check can give a random outcome
without removing any logical information. Measuring a Pauli operator that commutes
with all local checks but lies outside the check group removes one ebit.
For a single measured site $i$, let $\kappa_i=\kappa_{\{i\}}$. The single-site
estimate $p\sum_i\kappa_i$ counts individually exposed logical parities,
but joint readouts can contain redundant information or reveal parities
inaccessible to individual measurements. A Bell pair loses $2p-p^2$ ebits on average
because reading both ends repeats the same logical bit. Conversely, with local
check $X_1X_2$ and commuting noncheck $Z_1Z_2$, neither single-site
readout loses an ebit, whereas their joint parity does. These two
effects explain why the single-site estimate need not determine the
response at finite $p$.

\begin{figure}[!t]
\centering\figincl{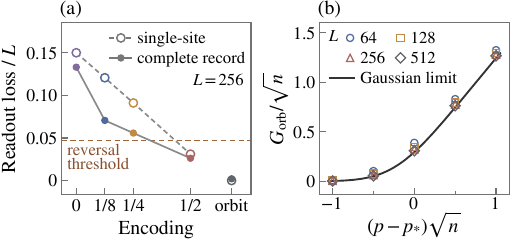}
\caption{Logical-information exposure and the dimensional bound.
(a) At $L=256,p=0.20$, open circles show the single-site estimate
$p\sum_i\kappa_i/L$ after the first gate layer. Filled circles show
the complete-record loss $G_d(p)/L$. The horizontal dashed line
$[G_{\rm ref}(p)+12]/256$ is the reversal threshold of
Eq.~\eqref{eq:ts-switch-main} for the initial gap of 12 ebits.
Intermediate-depth bars show pointwise 95\% confidence intervals.
The unencoded and orbit values are analytic.
(b) Uniform-orbit loss $G_{\rm orb}(p)/\sqrt n$ versus $(p-p_*)\sqrt n$,
where $n=L/2$, $m=3L/8$, $p_*=1/4$.
Exact recursions for $L=64,128,256,512$ approach the Gaussian
readout-count limit, Eq.~\eqref{eq:s11-closed}.}
\label{fig:full-response}
\end{figure}

Let $G_d(p)$ and $G_{\rm ref}(p)$ be the mean losses at the first readout,
including the preceding gate layer. Figure~\ref{fig:full-response}(a)
compares $G_d$ with the single-site estimate across encoding depths.
For the displayed finite
depths, joint readout loses less than that estimate. For the orbit,
the nonzero total response comes from joint parities
despite negligible single-site exposure.

\smallskip
\emph{Dimensional bound.}---After their outcomes are recorded, measured
qubits cannot carry entanglement. If $K_A$ and $K_B$ sites are read,
the surviving Schmidt states must fit on the more heavily measured
half, which has only $n-\max(K_A,K_B)$ unmeasured qubits. Readouts
beyond the redundancy $n-m$ therefore force a loss. Random encoding
nearly attains this dimensional limit. A fixed nonidentity Pauli operator
commutes with all checks of a uniform orbit with probability of order
$2^{-(n-m)}$, while the number of recorded Pauli products grows
exponentially with the readout size. Logical information typically
becomes accessible near the point where that size reaches the
redundancy, as in random monitored codes \cite{LeeYoshida2026Randomly}
and partially projected random states \cite{Sherry2026Information}.

\begin{theorem}[Asymptotic saturation of the dimensional bound]\label{thm:capacity}
For the uniform local Clifford orbit or independent exact unitary
two-design encoders of $\ket{\Phi_m^0}$, the mean first-readout loss
$G_\alpha(p)$, $\alpha\in\{\mathrm{orb},\mathrm{des}\}$, obeys
\begin{equation}
 G_\alpha(p)=\mathbb E\{[\max(K_A,K_B)-(n-m)]_+\}
 +O(n^{-1/2}),
 \label{eq:averaged-capacity-main}
\end{equation}
where $K_A,K_B\sim\mathrm{Bin}(n,p)$ are independent and
$[x]_+=\max(x,0)$. The remainder is nonnegative. The estimate is uniform for $0\le m\le n$ and for
$p$ in compact subintervals of $(0,1)$.
\end{theorem}

The bound on excess loss decays exponentially away from the lines
$K_A=n-m$, $K_B=n-m$, and $K_A=K_B$. Each binomial point mass is
$O(n^{-1/2})$, giving the remainder in Eq.~\eqref{eq:averaged-capacity-main} (End Matter~\ref{em:design} and
SM, Sec.~\ref{sec:up-capacity}).

At fixed initial entropy density, the dimensional bound implies
\begin{equation}
 G_\alpha(p)/L\longrightarrow\tfrac12[p-p_*]_+,
 \qquad p_*=1-m/n.
 \label{eq:up-threshold}
\end{equation}
Below $p_*$ by a fixed margin, a sharper tail bound makes the absolute
loss exponentially small. The threshold is set by the preparation's
redundancy, while $p_c$ characterizes the stationary dynamics.
Here $p_*=1/4$ exceeds the monitored transition at $p_c=0.159(3)$ (SM). The circuit at $p=0.20$ has an area-law stationary
state, yet its first readout leaves a fully scrambled preparation
almost intact. Fig.~\ref{fig:full-response}(b) shows how $G_{\mathrm{orb}}(p)$
varies with $p$ near $p_*$, where fluctuations in the number
of measurements produce an $O(n^{-1/2})$ threshold window.

For the unencoded state, the first gate layer acts on independent
blocks of two Bell pairs. A single readout removes one ebit from a
block, while reading opposite sides can reveal the same logical bit
twice. Averaging these finite-block records gives, for even $m,n$,
\begin{equation}
 G_0(p)=m\left[1-(1-p)^2\left(1+\frac{p^2}{5}\right)\right].
 \label{eq:d0-main}
\end{equation}
The horizontal line in Fig.~\ref{fig:full-response}(a) marks the
threshold for reversal at the first readout,
\begin{equation}
 D_d(1/2)>0\quad\Longleftrightarrow\quad G_d(p)>G_{\rm ref}(p)+(m-m').
 \label{eq:ts-switch-main}
\end{equation}
At $L=256$ and $p=1/5$, the uniform-orbit preparation
$\ket{\Phi_m^{\rm orb}}$ loses about half an ebit on average and the
reference about $10^{-3}$ ebit. The unencoded
state loses 34 ebits, nearly three times the initial gap of 12.
For two uniform Clifford orbits, the first readout preserves the mean
entropy order at every size and measurement rate:
$D_{\rm orb}(1/2)\le0$ (SM, Proposition~\ref{prop:shell-order}).

\smallskip
\emph{From readout to relaxation order.}---At $L=64$ and $d=L/2$, the
initial entropy order survives the first readout (SM, Table~\ref{tab:SM_shell_crossing}),
yet a later crossing is resolved [Fig.~\ref{fig:phenomenon-mechanism}(d)]. The second layer already
contains a gate across the cut, and later readouts can resolve logical
information that survived the first.

To follow the entanglement up to time $t$, fix the gates and measured
sites and average the outcomes with each input's Born probabilities.
Applying this circuit to halves of auxiliary Bell pairs gives a pure
Choi state of the output and the untouched partners for each allowed
record. The outcomes affect stabilizer signs but leave the Choi state's
entanglement unchanged across every cut.
The entropies of the two output halves and of one output half grouped
with either half of the untouched partner system give a common upper
bound on the mean output entropy of pure reference ensembles with mean
initial entropy at most $m$. Related cut bounds arise in random
tensor networks~\cite{Hayden2016Holographic,Nezami2020Multipartite}.
A two-copy average over the independent two-design encoders places the
encoded Bell-pair state's mean output entropy within two ebits of this
bound at every time (SM, Sec.~\ref{sec:SM_twodesign_alltime}).

\begin{theorem}[All-time control of macroscopic reversal]\label{thm:ordering}
Consider any preparation-independent, nonadaptive sequence of
Clifford gates and projective Pauli measurements, with outcomes averaged
according to each input's Born probabilities. Encode $\ket{\Phi_m^0}$
with independent exact unitary two-designs on the two halves to obtain
$\ket{\Phi_m^{\rm des}}$. Against any pure reference ensemble $\mathcal E$ with
mean initial entropy at most $m$,
\begin{equation}
 \overline S_{\mathcal E}(t)-\overline S_{\rm des}(t)
 \le2\qquad\text{for all }t.
 \label{eq:early-mean-order-main}
\end{equation}
For the random brickwork protocol specified above, with $L$ a
multiple of 64, $p=1/5$, $m=3L/8$ and $m'=21L/64$, $\ket{\Phi_m^0}$
against $\ket{\Phi_{m'}^{\rm orb}}$ satisfies
\begin{equation}
 D_0(1)\ge\frac{363}{6250}L-\frac25\sqrt L-\frac{18}{5}.
 \label{eq:switch-on-main}
\end{equation}
\end{theorem}

The first readout leaves the unencoded state with about $0.242L$ ebits
on average. The only gate across the cut can increase the half-chain
entropy by at most two ebits, and the next readout cannot increase the
mean entropy.
For the scrambled reference, replacing that gate by a suitable product
Clifford changes the final entropy by at most two ebits. Each half
undergoes $pL$ readouts on average over the cycle, each imposing at
most one new constraint. In this local circuit the reference comes
within a constant of the dimensional bound in mean entropy. At $p=1/5$ this gives
$\overline S_{\rm ref}(1)\ge 0.30L-O(\sqrt L)$, where the square-root term
allows for fluctuations in the readout counts. Using the exact first-readout loss from
Eq.~\eqref{eq:d0-main} in the subtraction gives the coefficient
$363/6250$ in Eq.~\eqref{eq:switch-on-main}. At these densities the
one-cycle reversal is extensive for every fixed $0.065<p<1/4$ at large
$L$ (SM, Sec.~\ref{sec:SM_analytic_switch}).

With the reference fixed, uniform encoding replaces the extensive
one-cycle reversal $D_0(1)=\Omega(L)$ with the all-time bound
$D_{\rm orb}(t)\le2$. The uniform-orbit curve in Fig.~\ref{fig:phenomenon-mechanism}(c) and its
peak in panel (d) are therefore bounded above by $2/L$.
The common output bound for the complete measurement history also gives
$D_0(t)\ge-2-O(Le^{-cL})$ for $t\ge1/2$
and some $c>0$ (SM, Sec.~\ref{sec:SM_no_return}). The unencoded state
therefore cannot regain an extensive lead in mean entanglement entropy
after the first readout. The one-cycle reversal and this subsequent bound
extend to exact two-design references, including Haar preparations
(SM, Sec.~\ref{sec:SM_analytic_switch}).

\smallskip
\emph{Quantum Mpemba effect.}---We diagnose entanglement relaxation
by the absolute deviation of the mean trajectory entropy from its
stationary value $S_\infty$, evaluated at the same cycle phase.
Suppose $m>m'>S_\infty$, so the more entangled preparation starts
farther from stationarity. If both means remain above $S_\infty$
when their order reverses, the lower one is closer. The two means
can also straddle $S_\infty$. In general, once $D_d(t)>0$, the
distance order has reversed precisely when
\begin{equation}
 S_\infty<\frac{\overline S_d(t)+\overline S_{\rm ref}(t)}2,
 \qquad t\in\mathbb N.
 \label{eq:strict-qme}
\end{equation}
For the pair in Theorem~\ref{thm:ordering},
the one-cycle midpoint is at least $3L/20-O(\sqrt L)$.
Consequently, any subextensive $S_\infty$, including an area-law
stationary entropy, gives the one-cycle QME for every sufficiently large multiple of 64.
When $S_\infty=o(L)$, the Wasserstein-1 distance to the stationary entropy
distribution likewise gives an extensive unencoded reversal at one cycle,
while exact two-design encoding excludes one at every integer cycle
(SM, Sec.~\ref{sec:SM_entropy_distribution}).

For finite $L$ and $0<p<1$, all pure inputs share a unique integer-cycle
stationary ensemble. Starting from the maximally mixed state
\cite{Gullans2020Dynamical,Magni2026Universal}, the monitored history
generates stabilizer checks that bound the entropy of every compatible
pure output. In particular, an input drawn from stationarity still
has mean output entropy $S_\infty$ after any whole number of
cycles. We can therefore bound $S_\infty$ using finite histories
(SM, Sec.~\ref{sec:up-stationary}). From 128 independent histories
per size, run for 32 cycles at $L=128$ and 64 cycles at $L=256$, we obtain
one-sided 99.5\% upper limits of $6.10$ and $8.74$ ebits, respectively.
The finite-size bounds in SM, Sec.~\ref{sec:SM_stationary_L256} give one-cycle
reversals above 1.25 and 7.66 ebits, with midpoints above 16.4 and
35.1 ebits. These bounds certify QME at both sizes.

\smallskip
\emph{Local implementation.}---The two-design encoders in
Theorem~\ref{thm:ordering} act on entire half-chains. What depth
suffices with nearest-neighbor gates?
Local random Clifford circuits also serve as encoders for approximate
quantum error correction
\cite{Gullans2021Quantum,Liu2026Approximate,Kroll2026Brickwork}.
For our contiguous Bell-pair seed, the product region initially holds
all local checks. Causality alone requires depth proportional to $L$
to spread them far enough for subextensive first-readout loss.
Independent local random Clifford circuits of sufficient linear depth
also give vanishing absolute loss below threshold and the all-time
bound $\overline S_{\mathcal E}(t)-\overline S_d(t)<4$ against any pure
reference ensemble $\mathcal E$ of mean initial entropy at most $m$.
The open-chain second moment \cite{Deneris2025Exact} supplies both
bounds at the same sufficient depth, asymptotically about $1.55L$
(SM, Sec.~\ref{sec:SM_encoding_depth}). Adding $O(\log L)$ layers
sharpens the all-time bound to $2+O(L^{-1})$, matching the
uniform-orbit bound asymptotically.

The random encoder gives a sharper necessity result. Beyond the support of all local checks, a single-site readout
reveals a logical bit. Under random Clifford gates, a check endpoint
follows a biased walk with drift $3/5$ per matching layer
\cite{Nahum2018Operator,vonKeyserlingk2018Operator}, slower than
the maximal causal speed. A macroscopic
interval therefore remains exposed until these fronts reach it.
For $m=3L/8$ and fixed $p>0$, vanishing loss density requires
$d\ge(5/8-o(1))L$ in this random ensemble
(SM, Sec.~\ref{sec:SM_random_front}).
The optimal depth for subextensive loss remains open.
The tested depth $d=L/2$ falls below the necessary threshold, so the
first-readout loss remains extensive, with
$\liminf_{L\to\infty}G_{L/2}(1/5)/L\ge 3/200$.
At all three tested sizes, the loss difference at this depth stays below
the initial gap $3L/64$, preserving the entropy order at the first readout
(SM, Table~\ref{tab:SM_shell_crossing}).

\smallskip
\emph{Discussion and outlook.}---Encoding depth offers experimental control over
relaxation while leaving the initial entanglement spectrum and the
monitoring sequence fixed. Superconducting processors have implemented
random circuits with mid-circuit readout \cite{Koh2023Measurement,Kamakari2025CrossEntropy}.
For ideal stabilizer trajectories the gates and measurement locations
determine the entropy. On a device, measuring the first-readout loss
and later entropy order would test whether the predicted suppression
survives imperfections. Correlations with classical simulations can
bound the mean trajectory entropy without postselection \cite{GarrattAltman2024}.
A crossing is certified when the reference's lower bound exceeds the
Bell-pair state's upper bound at the same time.

The two-ebit bound uses the flat Schmidt spectrum of the encoded
Bell-pair preparation. For i.i.d.\ nonflat pair spectra, independent
exact unitary two-design encoders still limit
the reference's mean entropy advantage to $O(\sqrt{L\log L})$ at all
times under Clifford gates and projective Pauli measurements
(SM, Sec.~\ref{sec:SM_twodesign_alltime}). In our $L=16$ simulations
with Haar-random gates, the unencoded state reverses the mean entropy
order within one cycle for all four spectra tested. Increasing the
encoding depth from zero to eight reduces the first-readout loss in
each case (SM, Fig.~\ref{fig:SM_nonclifford}). An all-time bound for non-Clifford dynamics remains open.

Each readout leaves the measured qubits in known product states. Local
gates within each half could reuse them as ancillas for the surviving
logical qubits. With feedforward~\cite{CaoEisert2026Shallow,CaoTangEisert2026AdaptiveFLO},
the encoder could choose which neighbors to couple to each ancilla.
Whether such adaptive circuits can preserve more entanglement at the next readout than fixed circuits of the same depth remains an open question.

\smallskip
\emph{Acknowledgments.}---This work was supported by
the Science Challenge Project (Grant No.~TZ2025017)
and the National Key Research and Development Program
of China (Grant Nos.~2021YFA1402104 and 2021YFA0718302).

\label{page:main-end}

\label{page:references-start}
\bibliography{bibliography}
\label{page:references-end}
% End Matter starts on a fresh page.
\clearpage
\twocolumngrid
\appendix
\begin{center}\textbf{\large End Matter}\end{center}
\setcounter{proposition}{0}
\renewcommand{\theproposition}{E\arabic{proposition}}
\renewcommand{\theHproposition}{EM.\arabic{proposition}}
\label{page:em-start}
\setcounter{section}{0}
\renewcommand{\thesection}{\Alph{section}}
\renewcommand{\theequation}{\Alph{section}\arabic{equation}}
\renewcommand{\theHsection}{EM.\Alph{section}}
\renewcommand{\theHequation}{EM.\Alph{section}.\arabic{equation}}
\refstepcounter{section}\setcounter{equation}{0}
\subsection*{A. What a complete record resolves}
\label{em:record}
For a pure input and projective readout on $J\subseteq A$, conditional
purity and $\sum_op_o\rho_B^o=\rho_B$ give
\begin{equation}
 \kappa_J=S(\rho_B)-\sum_op_oS(\rho_B^o)=I(O_J:B).
 \label{eq:holevo}
\end{equation}
For a stabilizer input with $k$ ebits, local signed checks support flat codes
\begin{equation}
 \begin{gathered}
 \rho_R=2^{-n_R}\sum_{g\in\mathcal K_R^+}g=2^{-k}\Pi_R,\\
 \dim\mathcal K_R=n_R-k.
 \end{gathered}\label{eq:half-code}
\end{equation}
Here $\mathcal K_R^+$ is the signed check group and $\mathcal K_R$
its binary label space. The orthogonal complement is symplectic. Let
$M=\operatorname{span}\{\bar Z_i:i\in J\}$ and
$P_o=\prod_{i\in J}\frac{I+(-1)^{o_i}Z_i}{2}$.
\begin{proposition}\label{prop:record}
Every allowed record loses the same number of ebits,
\begin{equation}
 \kappa_J=\dim(M\cap\mathcal K_A^\perp)-\dim(M\cap\mathcal K_A)
 \le k.
 \label{eq:record-rank}
\end{equation}
The conditional states on $B$ occupy $2^{\kappa_J}$ orthogonal equal-rank
sectors, each with total probability $2^{-\kappa_J}$.
\end{proposition}
The updated local check space is $M+(\mathcal K_A\cap M^\perp)$.
Rank-nullity for the commutation pairing shows that its dimension
increases by the difference in Eq.~\eqref{eq:record-rank}. The quotient
of commuting measured labels is isotropic in the $2k$-dimensional
logical space, so $\kappa_J\le k$.
Choose commuting Hermitian representatives $L_1,\ldots,L_{\kappa_J}$.
Their logical sectors have projectors
\begin{align}
 \Pi_x&=\Pi_A\prod_{\alpha=1}^{\kappa_J}
              \frac{I+(-1)^{x_\alpha}L_\alpha}{2},\notag\\
 \Pi_AP_o\Pi_A&=2^{\kappa_J}p_o\Pi_x,
 \label{eq:record-compressed-effect}
\end{align}
where the allowed record $o$ fixes the parities $x$. To obtain the
second line, expand $P_o$ in measured Pauli products. Terms that
anticommute with a check vanish between $\Pi_A$'s. The remaining
products impose the signed check constraints and the recorded logical
parities. Every nonidentity logical product has zero code trace, so
these projectors are orthogonal, sum to $\Pi_A$, and have rank
$2^{k-\kappa_J}$. Their trace fixes the factor $2^{\kappa_J}p_o$.

Taking traces and contracting in paired Schmidt bases, where
$\Pi_x^T$ acts on the $B$ code, gives
\begin{equation}
 p_o=2^{-|J|+\dim(M\cap\mathcal K_A)},\qquad
 \rho_B^o=\frac{\Pi_x^T}{2^{k-\kappa_J}}.
 \label{eq:record-sector-state}
\end{equation}
Since $\sum_op_o\rho_B^o=\Pi_B/2^k$, each sector has total weight
$2^{-\kappa_J}$. Measuring the orthogonal projectors $\Pi_x^T$ retrieves
$x$ perfectly. Outcomes within a sector give identical states on $B$,
so
\begin{equation}
 I_{\rm acc}(O_J:B)=I(O_J:B)=\kappa_J.
 \label{eq:record-equality}
\end{equation}
This equality uses orthogonal stabilizer sectors, whereas
Eq.~\eqref{eq:holevo} holds for every pure input. Conjugating by a local
Clifford extends the counting formula to any commuting Pauli readout.

\paragraph{Successive sides.}\label{em:successive-sides}
Conditioned on the $A$ record, the state remains pure and nonselective
measurement on $B$ preserves its $A$ marginal. Applying
Eq.~\eqref{eq:holevo} to these conditional states and averaging with
their Born weights gives the second loss $I(O_B:A\mid O_A)$. The two losses
therefore add to
\begin{equation}
 \kappa_{J_A,J_B}=I(O_A:B)+I(O_B:A\mid O_A).
 \label{eq:sequential-main}
\end{equation}
The first term is evaluated after $A$ readout and before $B$ readout.
The second uses the updated conditional state and its Born weights.

\refstepcounter{section}\setcounter{equation}{0}
\subsection*{B. Two-design dimensional bound with conditional Schmidt weights}
\label{em:design}
Let $\Lambda$ be the diagonal matrix of positive Schmidt weights of a
pure input, with rank $R\le N$ and $N\ge2$. An independent exact
unitary two-design on the $N$-dimensional measured side is followed by
a fixed complete projective readout with projectors $P_o$ of equal
rank $a\mid N$.
\begin{proposition}\label{prop:ts-instrument}
Its Born-averaged loss satisfies
\begin{equation}
 0\le\mathbb E_U\kappa\le\min\{S(\Lambda),\log_2(1+R/a)\}.
 \label{eq:ts-B}
\end{equation}
\end{proposition}
On the Schmidt support $\Pi$, put $Q_o=\Pi U^\dagger P_oU\Pi$ and
$\sigma_o=\sqrt\Lambda Q_o^T\sqrt\Lambda$, so
$p_o=\Tr\sigma_o$ and $\sum_o\sigma_o=\Lambda$.
For normalized $X,Y$ with
$\operatorname{supp}X\subseteq\operatorname{supp}Y$, write their
positive eigenpairs as $(x_i,\ket{i})$ and $(y_j,\ket{j})$.
Jensen's inequality with weights $x_i|\langle i\vert j\rangle|^2$ gives
\[
\begin{aligned}
 D(X\Vert Y)
 &=\sum_{ij}x_i|\langle i\vert j\rangle|^2\log_2(x_i/y_j)\\
 &\le\log_2\Tr(X^2Y^{-1}).
\end{aligned}
\]
For each $U$, the loss is
$\kappa=\sum_{o:p_o>0}p_oD(\sigma_o/p_o\Vert\Lambda)$.
Add the nonnegative quantity $\sum_op_o\log_2(Np_o/a)$, the classical
relative entropy to the uniform outcome law. Applying the preceding
inequality to each conditional state, then Jensen's inequality with
the Born weights, yields
\begin{align}
 \kappa
 &\le\sum_{o:p_o>0}p_o\log_2
       \frac{N\Tr(\sigma_o^2\Lambda^{-1})}{ap_o}\notag\\
 &\le\log_2\left[\frac Na\sum_o
       \Tr(\sigma_o^2\Lambda^{-1})\right],\notag\\
 \Tr(\sigma_o^2\Lambda^{-1})
 &=\Tr[\Lambda(Q_o^T)^2].
 \label{eq:ts-collision-support}
\end{align}
The inverse is taken on the Schmidt support and cancels before
averaging. With $F$ the swap, the standard two-copy twirl used in
decoupling \cite{Szehr2013Decoupling,Dupuis2014OneShot} gives
\begin{equation}
 \begin{gathered}
 \mathbb E(U^\dagger P_oU)^{\otimes2}=\alpha I+\beta F,\\
 N^2\alpha+N\beta=a^2,\qquad N\alpha+N^2\beta=a.
 \end{gathered}\label{eq:ts-twirl}
\end{equation}
The last two identities are its identity and swap traces. They give
$0\le\alpha\le a^2/N^2$ and $0\le\beta\le a/N^2$. Hence
$\mathbb E\Tr[\Lambda(Q_o^T)^2]=\alpha+R\beta$ and
\[
 \frac Na\sum_o\mathbb E\Tr[\Lambda(Q_o^T)^2]
 =\left(\frac Na\right)^2(\alpha+R\beta)\le1+R/a.
\]
Log concavity proves the upper bound. Equation~\eqref{eq:holevo}
gives $0\le\kappa\le S(\Lambda)$ after Born averaging.

\begin{samepage}
\begin{proposition}[Independent sides]\label{prop:ts-bilateral}
For independent exact two-design encoders on the two halves, followed
by fixed equal-rank projective readouts of retained ranks $a\le b$,
\begin{equation}
 \begin{gathered}
 0\le\mathbb E\kappa_{AB}\\
 \le\min\{S(\Lambda),
 \log_2[(1+R/a)(1+\min(R,a)/b)]\}.
 \end{gathered}
 \label{eq:ts-bilateral}
\end{equation}
\end{proposition}
\end{samepage}
Commute $U_B$ past the first readout. Its Born probability is unchanged
by unitaries on $B$, so conditioning on $U_A,o_A$ leaves the second
encoder's independent design law intact. The normalized conditional
state has rank at most $\min(R,a)$ and may be nonflat. Apply the
preceding bound to its spectrum, use monotonicity in rank, and sum
its Born weights. The two losses telescope by
Eq.~\eqref{eq:sequential-main}.

For a flat initial spectrum, every branch has Schmidt rank at most
$\min(R,a)$, so its loss is at least $[\log_2(R/a)]_+$.
Subtracting this dimensional minimum from Eq.~\eqref{eq:ts-bilateral}
gives
\begin{equation}
 \begin{aligned}
 0&\le\mathbb E\kappa_{AB}-[\log_2(R/a)]_+\\
 &\le\log_2\bigl[(1+\min\{R/a,a/R\})\\
 &\qquad\quad\times(1+\min\{R,a\}/b)\bigr]\le2.
 \end{aligned}
 \label{eq:ts-flat}
\end{equation}
Each factor inside the logarithm is at most two because $a\le b$.
Uniform Cliffords and Haar encoders meet the two-design hypothesis
\cite{Dankert2009Exact}. Finite brickwork depth is treated in SM,
Sec.~\ref{sec:SM_encoding_depth}.
\label{page:em-end}
\clearpage
\onecolumngrid
\label{page:sm-start}
\begin{center}
\textbf{\large Supplemental Material for\\
``Quantum Mpemba effect from logical-information exposure in monitored circuits''}
\end{center}
\renewcommand{\thetable}{S\arabic{table}}
\renewcommand{\thefigure}{S\arabic{figure}}
\renewcommand{\theequation}{S\arabic{equation}}
\renewcommand{\theHtable}{S\arabic{table}}
\renewcommand{\theHfigure}{S\arabic{figure}}
\renewcommand{\theHequation}{S\arabic{equation}}
\setcounter{equation}{0}
\setcounter{figure}{0}
\setcounter{table}{0}
\setcounter{section}{0}
\renewcommand{\thesection}{S\arabic{section}}
\setcounter{secnumdepth}{3}
\renewcommand{\appendixname}{Section}
\renewcommand{\thesubsection}{\thesection.\Alph{subsection}}
\renewcommand{\thesubsubsection}{\thesubsection.\roman{subsubsection}}
\makeatletter
\renewcommand{\p@subsection}{}
\renewcommand{\p@subsubsection}{}
\makeatother
\renewcommand{\theHsection}{SM.\arabic{section}}
\renewcommand{\theHsubsection}{SM.\arabic{section}.\arabic{subsection}}
\renewcommand{\theHsubsubsection}{SM.\arabic{section}.\arabic{subsection}.\arabic{subsubsection}}
\counterwithout*{equation}{section}

\medskip

%======================================================================

\renewcommand{\theproposition}{S\arabic{proposition}}
\renewcommand{\theHproposition}{SM.\arabic{proposition}}
\setcounter{proposition}{0}

\section{Single-site exposure and the first gate layer}
\label{sec:SM_analytical}
For a pure stabilizer state and a site $i$ in half $R$,
\begin{equation}
 \kappa_i=\mathbf1[\bar Z_i\in\mathcal K_R^\perp
                    \setminus\mathcal K_R],\qquad
 \chi=\sum_{i=1}^L\kappa_i .\label{eq:kappa}
\end{equation}
A check fixes the outcome; an anticommuting check gives
$\Pi_R Z_i\Pi_R=0$; a commuting noncheck resolves two equal logical
sectors, losing one ebit. For arbitrary pure states, the Holevo identity
gives $0\le\kappa_i\le H(O_i)\le1$. For an independent Bernoulli-$p$
mask, $\mathcal G_p=\mathbb E_J\kappa_J$ has the fixed-$L$ expansion
\begin{equation}
 \mathcal G_p=p\chi+p^2\sum_{i<j}(\kappa_{ij}-\kappa_i-\kappa_j)+O(p^3).
 \label{eq:linear}
\end{equation}
Bare Bell pairs give $\mathcal G_p(\ket{\Phi_m^0})=m(2p-p^2)$ and
$\chi(\ket{\Phi_m^0})=2m$. The uniform signed stabilizer orbit at entropy
$m$ gives, by Eq.~\eqref{eq:shell-transition},
\begin{equation}
 \mathbb E_{\rm orb}\chi/L=q(n,m)
 =2^{-(n-m)}\frac{1-2^{-2m}}{1-2^{-2n}}.\label{eq:random-em}
\end{equation}
Uniform orbits are invariant under the intra-half first gates, whereas
for a finite-depth input the first gate layer can change $\chi$ before readout.

\paragraph{Gate-first response.}\label{sec:SM_unencoded_details}
\begin{proposition}\label{prop:d0-quartic}
For even $m,n=L/2$, Bell pairs $(a,L+1-a)$ with $a\le m$, and
$\ket{0}$ qubits elsewhere, independent uniform two-qubit
Cliffords on $(1,2),(3,4),\ldots$ give
\begin{equation}
 G_0(p)=\frac m2
 \left(4p-\frac{12}{5}p^2+\frac45p^3-\frac25p^4\right).
 \label{eq:d0-quartic}
\end{equation}
\end{proposition}
No bond crosses the cut or active-region boundary, leaving $m/2$
independent blocks, each formed from two Bell pairs. Measuring one site
in a block removes one ebit; measuring two sites on the same side, or
any three or four sites, removes two.
After the first readout on one side, the opposite code has one check uniform
among 15 nonidentity Pauli operators; six make the next measured Pauli operator a
commuting noncheck. Thus measuring one site on each side removes
$1+6/15=7/5$ ebits on average;
the size-two sum is $48/5$, and the block loss is
$4p(1-p)^3+(48/5)p^2(1-p)^2+8p^3(1-p)+2p^4$, proving the formula.

\section{Dimensional bound and the counting window}
\label{sec:up-capacity}
\paragraph{Conditional orbit law.}
Let $\Omega(a,b,e)$ be the uniform pure-stabilizer ensemble on $a:b$
qubits at entropy $e$, with all signs and modulo global phase. Local
Cliffords map every member to $e$ Bell pairs and product checks, and
local Paulis fix the signs, proving transitivity
\cite{Fattal2004Entanglement}.
\begin{proposition}\label{prop:shell-closure}
Measuring and removing a specified $A$ qubit, conditional on a legal
outcome and residual entropy $e'$, leaves $\Omega(a-1,b,e')$ uniform.
\end{proposition}
A residual local Clifford extended by identity on the measured site
commutes with the projector and preserves the input law and Born
weight. The conditional output law is invariant under this transitive
group, also after any preceding outcome/entropy history.

The local check space and centralizer have $2^{a-e}$ and $2^{a+e}$
labels. Counting centralizer nonchecks among $2^{2a}-1$ nonzero labels
gives the decrement probability $q(a,e)$ and, from $\pi_0=\delta_m$,
the entropy law after $j$ readouts:
\begin{align}
 q(a,e)&=\frac{2^{a+e}-2^{a-e}}{2^{2a}-1}
       =2^{e-a}\frac{1-2^{-2e}}{1-2^{-2a}},\label{eq:shell-transition}\\
 \pi_{j+1}(x)&=\pi_j(x)[1-q(n-j,x)]
                   +\pi_j(x+1)q(n-j,x+1).\label{eq:shell-distribution}
\end{align}
For $a\ge1$, $q(a,0)=0$, $q(a,a)=1$; each readout reduces $a$ by one,
and none occurs at $a=0$. Unsupported terms vanish. If $T_n(k)$ is the
$k$-step transition, permutation invariance and independent mask counts give
\begin{equation}
 K_n(p)=\sum_{k=0}^n\binom nk p^k(1-p)^{n-k}T_n(k),\qquad
 \mu_f=\delta_mK_{n_A}(p_A)K_{n_B}(p_B),\quad
 G_{\rm orb}=\sum_x(m-x)\mu_f(x).\label{eq:shell-full}
\end{equation}
The second side receives the full first-side entropy distribution.
Since the projectors commute, measuring either side first gives the
same final distribution.
\begin{proposition}\label{prop:shell-order}
At fixed side sizes and counts, increasing initial orbit entropy from
$m_1$ to $m_2$ stochastically increases terminal entropy, with mean
increase in $[0,m_2-m_1]$.
\end{proposition}
Couple decrements by one uniform variable. Since
$q(a,e+1)-q(a,e)=(2^{a+e}+2^{a-e-1})/(2^{2a}-1)>0$, the higher process
decrements whenever the lower does. Their integer gap cannot grow or
cross zero, and equal entropies stay equal. Conditional closure extends
the coupling through both sides while preserving each signed-ensemble
and Born marginal, even for different physical signed records.

\paragraph{Fixed-count remainders.}\label{sec:SM_capacity_remainder}
For total loss $\ell$ at counts $(k_A,k_B)$, write
$g_{n,m}=\mathbb E\ell$, $\delta=n-m$, $K=\max(k_A,k_B)$,
$h=\min(k_A,k_B)$, and $b=[K-\delta]_+$. For the uniform Clifford orbit
we first show
\begin{equation}
 0\le g_{n,m}(k_A,k_B)-[\max(k_A,k_B)-(n-m)]_+\le\tfrac32.
 \label{eq:up-capacity}
\end{equation}
After loss $T$ and $j$ readouts on the current side, the next decrement
$I$ obeys $\mathbb E(2^{T+I}\mid\mathcal H)
=2^T[1+q(n-j,m-T)]\le2^T+2^{j-\delta}$.
Summing over both sides without resetting $T$ gives
\begin{equation}
 \mathbb E2^\ell\le Q=1+2^{-\delta}(2^{k_A}+2^{k_B}-2).
 \label{eq:SM_shell_moment}
\end{equation}
Retained dimension gives $\ell\ge b$ on every branch, so the integer
$\eta=\ell-b\ge0$ satisfies $\mathbb E2^\eta\le2^{-b}Q\le3$: each of the
three positive terms is at most one since $b\ge0$ and
$k_R-\delta-b\le0$. Using $\eta\le2^{\eta-1}$ proves
Eq.~\eqref{eq:up-capacity}. Also using $\eta\le m-b$ and $\eta\le2^\eta-1$
gives the sharper orbit bound below. For two-design encoders, substitute
$R=2^m$, $a=2^{n-K}$, and second-side retained rank $2^{n-h}$ in
Eq.~\eqref{eq:ts-flat} of End Matter~\ref{em:design}; since
$\min(R,a)=2^{m-b}$, the fixed-count bounds are
\begin{align}
 0\le g_{n,m}(k_A,k_B)-b
 &\le\min\{m-b,\,2^{-b}Q-1,\,2^{-b-1}Q\},
 \label{eq:capacity-count-shell}\\
 0\le g^{\rm des}_{n,m}(k_A,k_B)-b
 &\le\min\bigl\{m-b,\,
 \log_2[(1+2^{-|K-\delta|})(1+2^{h-\delta-b})]\bigr\}.
 \label{eq:capacity-count-design}
\end{align}
Binomial averaging at $n=128$, $p=1/5$ bounds $(G_{\rm orb},G_{\rm des})$
by $(0.541,0.578)$ ebits at $m=96$ and $(0.00119,0.00152)$ at
$m=84$, rounded upward.

For independent $K_A,K_B\sim\operatorname{Bin}(n,p)$ and
$M_n(p)=\max_k\Pr(K_A=k)$,
\begin{align}
 0\le G_{\rm orb}-
 \mathbb E[\max(K_A,K_B)-\delta]_+&\le9M_n(p),\notag\\
 0\le G_{\rm des}-
 \mathbb E[\max(K_A,K_B)-\delta]_+&\le\frac{9M_n(p)}{\ln2}.
 \label{eq:capacity-averaged-remainder}
\end{align}
Indeed, separating $K\le\delta$ and $K>\delta$ in $2^{-b}Q-1$ bounds
the orbit remainder by
$2^{-|k_A-\delta|}+2^{-|k_B-\delta|}+2^{-|k_A-k_B|}$;
$\ln(1+u)\le u$ gives the same two-design bound divided by $\ln2$.
Each term averages to at most $3M_n(p)$: use
$\sum_{j\in\mathbb Z}2^{-|j|}=3$ and the fact that $K_A-K_B$ also has
point masses at most $M_n(p)$. On compact subintervals of $0<p<1$,
$M_n(p)=O(n^{-1/2})$ uniformly in $p$, so the absolute remainder is
$O(n^{-1/2})$ uniformly there, even in the counting window. This uses flat Schmidt weights; the nonflat
bound in End Matter~\ref{em:design} depends on Schmidt rank.

For the Clifford orbit, the integer deficit $\eta$ gives
$\Pr(\ell\ne b)\le\mathbb E\eta\le9M_n(p)$.
The dimensional bound is therefore attained exactly with probability
$1-O(n^{-1/2})$ at fixed $0<p<1$.

\paragraph{Basic absolute protection.}\label{sec:ts-window}
Choose $p<q<\delta/n$. On $K_A,K_B\le\lfloor nq\rfloor$, apply
$\ell\le2^\ell-1$ to Eq.~\eqref{eq:SM_shell_moment}; on its complement,
use $\ell\le m$ and the binomial Chernoff bound:
\begin{align}
 G_{\rm orb}(p)&\le2^{1-\delta}(2^{\lfloor nq\rfloor}-1)
                       +2m e^{-n\operatorname{kl}(q\Vert p)},\label{eq:s11-exp}\\
 G_{\rm des}(p)&\le\frac2{\ln2}2^{-\delta+\lfloor nq\rfloor}
                       +2m e^{-n\operatorname{kl}(q\Vert p)}.\label{eq:ts-rank-protection}
\end{align}
Here $\operatorname{kl}(q\Vert p)=q\ln(q/p)+(1-q)\ln[(1-q)/(1-p)]$.
For the second bound, End Matter~\ref{em:design} bounds each half's
conditional projection loss by $2^{-\delta+\lfloor nq\rfloor}/\ln2$ on
the count event. This also holds for arbitrary nonzero Schmidt weights
with $m=\log_2R$, $\delta=n-m$, since initial entropy is at most $m$.
A fixed positive redundancy gap gives exponentially vanishing loss;
however, at $L=256$, $m=96$, $p=1/5$, $q=9/40$, the tail term is about
151 ebits. Finite-size orbit values use Eq.~\eqref{eq:shell-full}.

\paragraph{Macroscopic law and Gaussian window.}
Equation~\eqref{eq:capacity-averaged-remainder} and the law of large
numbers give Eq.~\eqref{eq:up-threshold} for $m/L\to s\in(0,1/2)$,
$p_*=1-2s$. For $m=2sn+O(1)$ and $p_n=p_*+x/\sqrt n$, the centered
binomials divided by $\sqrt n$ converge to independent normals of
variance $\sigma^2=p_*(1-p_*)$. Bounded second moments make their
positive-part maximum uniformly integrable. With $v=x/\sigma$ and
standard normal CDF and density $\Phi,\varphi$, this gives
\begin{align}
 \frac{G_\alpha(p_n)}{\sqrt n}\longrightarrow F_\sigma(x)
 &=\int_0^\infty[1-\Phi((t-x)/\sigma)^2]dt,\label{eq:s11-gaussian}\\
 F_\sigma(x)&=x[1-\Phi(-v)^2]+2\sigma\varphi(v)\Phi(-v)
                 +\frac\sigma{\sqrt\pi}\Phi(\sqrt2v).
 \label{eq:s11-closed}
\end{align}
The closed form integrates the maximum's density $2\varphi(z)\Phi(z)$.
Fixed counts $k_A=k_B=\lfloor np_*+x\sqrt n\rfloor$ instead give
$g=\sqrt n[x]_++O(1)$, identifying count fluctuations as the source of
the window. Figure~\ref{fig:full-response}(b) compares $F_\sigma(x)$ with
the 20 $s=3/8$ recursion points.

\section{Finite local depth: necessity and protection}
\label{sec:SM_encoding_depth}
Let $L=2n$, with even $n\ge8$. Independent half-chain encoders use $d$
alternating matchings $O=\{(1,2),(3,4),\ldots\}$ and
$E=\{(2,3),(4,5),\ldots\}$, starting with $O$, with iid uniform
two-qubit Cliffords. An independent physical $O$ layer $U_0$ precedes
readout. In reflected Bell-partner coordinates, the seed has logical
sites $1,\ldots,m$, checks
$C_0=\operatorname{span}\{Z_{m+1},\ldots,Z_n\}$, and $\delta=n-m$.

\subsection{Geometric necessity}
For even $0<m<n$, the first $O$ layer cannot extend checks into the
logical interval; each later matching advances their boundary by at
most one site. Including $U_0$, the range is $2\lfloor d/2\rfloor$:
the encoder contributes at most $d-1$ for $d\ge1$, and $U_0$ contributes
one only after an $E$ layer. Thus at least
$[m-2\lfloor d/2\rfloor]_+$ sites per half lie outside all check
supports. Their measured-$Z$ span centralizes and intersects the checks
trivially. Reading them first loses one ebit per selected site by
Eq.~\eqref{eq:record-rank}; later local readouts have nonnegative mean
loss. Counting one half gives
\begin{equation}
 G_d(p)\ge p[m-2\lfloor d/2\rfloor]_+.
 \label{eq:depth-geometric}
\end{equation}
For $p_L\ge p_0>0$ and $m/L\to s>0$, subextensive loss therefore requires
$\liminf d_L/L\ge s$. At $m=0$ the loss vanishes; at $m=n$ no checks
exist and $G_d\ge pn$. At odd depth, the last encoder $O$ layer is
absorbed by uniform $U_0$, giving $G_{2k+1}=G_{2k}$.

\subsection{A sharper lower bound for random encoders}
\label{sec:SM_random_front}
For fixed $p>0$, $m/L\to s\in(0,1/2)$ and $d/L\to r$, uniform random
Clifford gates sharpen the deterministic light-cone bound to
\begin{equation}
 \liminf_{L\to\infty}\frac{G_d(p)}L\ge p[s-3r/5]_+.
 \label{eq:depth-random-front}
\end{equation}
Hence subextensive loss requires $r\ge5s/3$ in this ensemble.
After absorbing the last $O$ layer when $d$ is odd, each initial check
$Z_j$, $j>m$, evolves through $T=2\lfloor d/2\rfloor+1$ alternating
matchings. A random two-qubit Clifford sends a nonzero Pauli label uniformly
to 15 labels, with support $10$, $01$, $11$ of probabilities
$1/5$, $1/5$, $3/5$~\cite{Nahum2018Operator,vonKeyserlingk2018Operator}.
For its left endpoint $X_i$, let $B_i$ indicate an outward next matching.
Away from the physical ends, $B_{i+1}$ is fresh Bernoulli-$4/5$ and
$X_i-X_{i+1}=B_i+B_{i+1}-1$. Thus
$X_0-X_u=\sum_{i=1}^u(2B_i-1)+B_0-B_u$, with drift $3/5$ per matching.

Fix $1\le q<m$. Each excursion entering the logical block at $m$,
stopped on return to $m+1$ or first hit of $q$, has only internal
updates. The maximal exponential bound for this walk gives hitting
probability at most
$e^{\lambda-\lambda(m-q)}[(4e^\lambda+e^{-\lambda})/5]^T$ for $\lambda>0$.
A union bound over all generators and at most $T$ entrances per
generator gives
\begin{equation}
 \Pr\{\text{some check reaches }[1,q]\}\le\epsilon_{q,T}
 :=\min\left\{1,(n-m)T e^{\lambda-\lambda(m-q)}
 \left(\frac{4e^\lambda+e^{-\lambda}}5\right)^T\right\}.
 \label{eq:depth-front-tail}
\end{equation}
On the complementary
event, all checks act trivially on $[1,q]$; reading that
interval first therefore gives
\begin{equation}
 G_d(p)\ge pq(1-\epsilon_{q,T}).
 \label{eq:depth-front-loss}
\end{equation}
For $v=(m-q)/T\in(3/5,1)$, choosing
$\lambda=\frac12\ln[(1+v)/(4(1-v))]$ gives a polynomial prefactor times
$e^{-T\operatorname{kl}((1+v)/2\Vert4/5)}$.
Taking $q/L\uparrow s-3r/5$ proves Eq.~\eqref{eq:depth-random-front};
$r=0$ follows from Eq.~\eqref{eq:depth-geometric}.

\subsection{Protection for a fixed monitored history}
Replace the cross-cut gates of a preparation-independent nonadaptive
history $\omega$ of Clifford gates and projective Pauli measurements by the local Cliffords of
Sec.~\ref{sec:SM_boundary_cost}. On half $R$, write $z_R$ for the
input-check dimension, $H_R=n-z_R$ for the logarithmic branch rank,
and $M_\omega^{\rm cut}=\min(m,H_A,H_B)$. Define the nonnegative deficit
\begin{equation}
 a_d(\omega,t)=M_\omega^{\rm cut}-\mathbb E_{\rm enc,Born}s_d^{\rm cut}(t),
 \qquad\varepsilon_d(L,t)=\mathbb E_\omega a_d(\omega,t),
 \label{eq:depth-defect}
\end{equation}
where $s_d^{\rm cut}$ is the trajectory entropy in this local protocol.
Reversing the adjoint instrument on $I$ gives the same unsigned
input-check space $W_R$ on every nonzero branch: Clifford conjugation
and $F_P(C)=(C\cap P^\perp)+\langle P\rangle$ ignore legal signs.
The normal form~\eqref{eq:SM_history_normal_form} is
$K_{R,o}=c_{R,o}C_{R,o}\Pi_{W_R,s(o)}$, with output Clifford $C_{R,o}$
and $\dim W_R=z_R$. For fixed stabilizer input, all legal static
readouts of $W_A\oplus W_B$ have one entropy $s_W$; output local
Cliffords preserve it. Completeness therefore gives
$\sum_o w_\psi(o)s_{\psi,o}^{\rm cut}=s_W$, reducing the Born mean to
static readout of these spaces.

\paragraph{Second-moment contraction.}
The second-moment formulas below also hold for even $n\ge4$; for $n=2$, one
uniform two-qubit Clifford already gives the uniform half-chain ensemble.
Let $P_O,P_E$ be the orthogonal two-copy matching twirls and $\Pi$ the
global identity/swap projector. The local Cliffords have Haar second
moments and generate the global Clifford group, so the common range
is $\operatorname{ran}\Pi$.
Ref.~\cite[Theorem~1, Eq.~(11)]{Deneris2025Exact} gives the largest
nontrivial eigenvalue of $P_OP_E$:
\begin{equation}
 \lambda_n=c_n^2,\qquad c_n=\frac45\cos(\pi/n).
 \label{eq:depth-open-spectrum}
\end{equation}
Its layer comprises two matchings; our depth counts each separately.
Put $O_0=P_O-\Pi$, $E_0=P_E-\Pi$, $A_0=O_0E_0O_0\ge0$.
Since $\|A_0\|=c_n^2$,
$[(E_0O_0)^k]^\dagger(E_0O_0)^k=A_0^{2k-1}$ and
$O_0(E_0O_0)^k=A_0^k$ for $k\ge1$ give
\begin{equation}
 \|\mathcal M_d-\Pi\|_{2\to2}=\beta_{n,d}:=
 \begin{cases}1,&d=0,\\c_n^{d-1},&d\ge1,\end{cases}
 \quad\mathcal M_{2k}=(P_EP_O)^k,\quad
 \mathcal M_{2k+1}=P_O(P_EP_O)^k.
 \label{eq:depth-beta}
\end{equation}
Restricting the adjoint twirl to the invariant span of the orthonormal
nonzero-Pauli vectors $(P_q\otimes P_q)/2^n$ yields, for $N=4^n-1$,
\begin{equation}
 (P_d)_{rq}=\Pr(U_d^\dagger P_qU_d=\pm P_r),\qquad
 \left\|P_d-\frac{\mathbf1\mathbf1^T}{N}\right\|_{2\to2}\le\beta_{n,d}.
 \label{eq:depth-pauli-norm}
\end{equation}
The bound is unchanged by adjoints or reversed matchings.

\paragraph{Complete readout and rank deficit.}
For a check space $C$ of dimension $n-e$ and readout space $W$ of
dimension $k$, Eq.~\eqref{eq:record-rank} gives
$2^X=|W\cap(UC)^\perp|/|W\cap UC|$.
Bounding the denominator by one and applying
Eq.~\eqref{eq:depth-pauli-norm} to the nonzero-label indicators of
$C^\perp,W$ gives, by Cauchy--Schwarz,
\begin{equation}
 \mathbb E2^X\le1+2^{e-n}(2^k-1)
 +\beta\sqrt{(2^{n+e}-1)(2^k-1)},\qquad\beta=\beta_{n,d}.
 \label{eq:depth-one-sided-moment}
\end{equation}
Put $k_R=z_R$, $A_R=2^{-\delta}(2^{k_R}-1)$, $x=2^n\beta$,
and $u_R=1+A_R+x\sqrt{A_R}$. Read $A$ first, with loss $X$.
Its outcome probability is independent of $U_B$, preserving the
second encoder's conditional law. The partner code has entropy $m-X$,
so its loss $Y$ obeys
$\mathbb E(2^Y\mid U_A,o_A)\le1+A_B2^{-X}+x\sqrt{A_B}\,2^{-X/2}$.
Using $\mathbb E2^{X/2}\le\sqrt{\mathbb E2^X}\le\sqrt{u_A}$, and
minimizing over the commuting readout orders with their Born weights,
\begin{equation}
 \mathbb E2^\ell\le\mathcal U:=\min\{
 u_A+A_B+x\sqrt{A_Bu_A},\ u_B+A_A+x\sqrt{A_Au_B}\},
 \qquad\ell=X+Y.
 \label{eq:depth-two-sided-moment}
\end{equation}
For $b=[\max(k_A,k_B)-\delta]_+$, the rank deficit is the nonnegative
integer $\eta=\ell-b=M_\omega^{\rm cut}-s^{\rm cut}$.
Since $\eta\le1+2^{\eta-2}$, $a_d\le\min\{m,1+R/4\}$ with
$R=2^{-b}\mathcal U$. Using $A_R\le2^b$ bounds its constant and two
error terms by $3$, $x$, $x\sqrt{2+x}$, respectively:
\begin{align}
 R&\le R_*(x)=3+x(1+\sqrt{2+x}),\quad
 a_d\le B_{\rm enc}(L,d):=\min\{m,1+R_*(x)/4\},
 \label{eq:depth-uniform-moment}\\
 d&\ge1+\left\lceil\frac{n\ln2}{-\ln c_n}\right\rceil
 \quad\Longrightarrow\quad a_d,\varepsilon_d\le2+\frac{\sqrt3}{4}.
 \label{eq:depth-linear-family}
\end{align}
This holds uniformly in the history, with sufficient ratio
$d/L\to\ln2/[2\ln(5/4)]\simeq1.55$.

For first readout, fix the local layer $U_0$ and absorb it into the
readout directions. Choose $0<p<r<\delta/n$ and put
$h=\lfloor nr\rfloor$, $a_h=2^{-\delta}(2^h-1)$, $t_h=x\sqrt{a_h}$.
Independent counts $K_R\sim\operatorname{Bin}(n,p)$ obey
$\Pr(K_R>h)\le e^{-n\operatorname{kl}(r\Vert p)}$.
Use $\ell\le2^\ell-1$ in Eq.~\eqref{eq:depth-two-sided-moment} when
both counts are at most $h$, and $\ell\le m$ otherwise:
\begin{equation}
 G_d(p)\le2a_h+t_h(1+\sqrt{1+a_h+t_h})
 +2m e^{-n\operatorname{kl}(r\Vert p)}.
 \label{eq:depth-first-tail}
\end{equation}
For fixed positive gaps $r-p$, $\delta/n-r$ and $x=O(1)$ this is
exponentially small up to $m$. At $m/L=3/8$, $p=1/5$, $r=9/40$,
$a_h\le2^{-n/40}$ and $t_h\le x2^{-n/80}$, giving $G_d\to0$ at the
stated depth; the finite-size tail limitation is as in
Sec.~\ref{sec:ts-window}.

\subsection{Finite-depth protection for the complete history}
\label{sec:SM_depth_alltime}
Let $S_0$ be the unsigned stabilizer space of a pure stabilizer seed
of entropy $m$, and $S_d=(U_A\oplus U_B)S_0$ its encoding by independent
$n$-qubit Cliffords. For any preparation-independent nonadaptive
history of Clifford gates and projective Pauli measurements, use $W\subseteq F\subseteq W^\perp$ and
$f,f_A,f_B,b$ of Sec.~\ref{sec:SM_full_history}, where
\begin{equation}
 f\le2n+b,\qquad f_A,f_B\le n+m+b.
 \label{eq:depth-alltime-dimensions}
\end{equation}
Every realization has $M_\omega-s_d=q(S_d)-b\ge0$.
Write the forward nonzero-Pauli transitions as $P_R=J+E_R$, with
$K=4^n-1$, $J=\mathbf1\mathbf1^T/K$ and
$\|E_R\|_{2\to2}\le\beta=\beta_{n,d}$ by
Eq.~\eqref{eq:depth-pauli-norm}. Put $x=2^n\beta$.
We compare $\mathbb E|S_d\cap F|$ with its uniform-orbit value.

Let $u_R,v_R$ indicate the nonzero labels of $S_0\cap R,F\cap R$;
then $\dim(S_0\cap R)=n-m$. For labels nonzero on both sides, put
$H_{ab}=\mathbf1_{(a,b)\in S_0}$, $G_{ab}=\mathbf1_{(a,b)\in F}$.
Their expected contribution is $\langle G,P_AHP_B^T\rangle_F$, where
$\langle X,Y\rangle_F=\Tr(X^TY)$ and
\begin{equation}
 P_AHP_B^T=JHJ+E_AHJ+JHE_B^T+E_AHE_B^T.
 \label{eq:depth-alltime-expansion}
\end{equation}
The first term is uniform. Each nonempty fiber of $F\to A$ has
$2^{f_B}$ labels before removing zero on $B$, and at most $2^{f-f_B}$
fibers exist. The same count for $S_0$, whose dimension is $2n$ and
$B$-only dimension is $n-m$, bounds the row sums:
\begin{equation}
 \rho_G=G\mathbf1,\quad\rho_H=H\mathbf1,\qquad
 \|\rho_G\|_2\le2^{(f+f_B)/2},\quad
 \|\rho_H\|_2\le2^{(3n-m)/2}.
 \label{eq:depth-alltime-fibers}
\end{equation}
Cauchy--Schwarz, the dimension bounds~\eqref{eq:depth-alltime-dimensions},
and $\|H\|_F\le2^n$, $\|G\|_F\le2^{f/2}$ now give
\begin{align}
 |v_R^TE_Ru_R|&\le\beta\sqrt{(2^{f_R}-1)(2^{n-m}-1)}
 \le\beta2^{n+b/2}\le2^b x,
 \label{eq:depth-alltime-local-error}\\
 |\langle G,E_AHJ\rangle_F|
 &=\frac{|\rho_G^TE_A\rho_H|}{K}
 \le\frac{\beta2^{(f+f_B+3n-m)/2}}K
 \le\frac{2^b x}{1-4^{-n}},
 \label{eq:depth-alltime-mixed-error}\\
 |\langle G,E_AHE_B^T\rangle_F|
 &\le\beta^2\|G\|_F\|H\|_F\le\beta^2 2^{2n+b/2}\le2^b x^2.
 \label{eq:depth-alltime-quadratic-error}
\end{align}
There are two local terms; column sums give the same mixed bound for
$JHE_B^T$. Adding these five errors to the uniform value $\le4\,2^b$
from Eq.~\eqref{eq:SM_full_count_bound}, and using $|S_d\cap W|\ge1$,
\begin{align}
 2^{-b}\mathbb E|S_d\cap F|&\le C_n(x),\qquad
 C_n(x)=4+2x+\frac{2x}{1-4^{-n}}+x^2,
 \label{eq:depth-alltime-C}\\
 \mathbb E2^{\eta_d}
 &=2^{-b}\mathbb E\frac{|S_d\cap F|}{|S_d\cap W|}\le C_n(x),
 \qquad\eta_d=M_\omega-s_d\ge0.
 \label{eq:depth-alltime-moment}
\end{align}
By Eq.~\eqref{eq:SM_design_common_cap} and the cap identity in
Sec.~\ref{sec:SM_twodesign_alltime}, any pure reference ensemble $\mathcal E$
of mean initial entropy at most $m$ satisfies
$\overline S_{\mathcal E}\le M_\omega$. Jensen's inequality and averaging
over the preparation-independent history give
\begin{equation}
 \overline S_{\mathcal E}(t)-\overline S_d(t)
 \le\log_2 C_n(2^n\beta_{n,d})\qquad\text{for every }t.
 \label{eq:depth-alltime-order}
\end{equation}
For each encoded stabilizer input and fixed schedule, all legal
outcomes have identical unsigned stabilizer spaces, so the encoder
averages retain the Born weights. The depth in Eq.~\eqref{eq:depth-linear-family}
ensures $x=2^n c_n^{d-1}\le1$, hence
\begin{equation}
 \overline S_{\mathcal E}(t)-\overline S_d(t)
 \le\log_2\left(7+\frac2{1-4^{-n}}\right)<4
 \qquad\text{for every }t,
 \label{eq:depth-alltime-four}
\end{equation}
as $7+2/(1-4^{-n})\le29/3<16$ for $n\ge1$. The sufficient ratio
remains $d/L\to\ln2/[2\ln(5/4)]\simeq1.55$, with first-readout
protection still given by Eq.~\eqref{eq:depth-first-tail}.
Increasing the sufficient depth in Eq.~\eqref{eq:depth-linear-family}
by $\lceil\ln L/\ln(5/4)\rceil$ gives $x\le L^{-1}$ and hence
$\overline S_{\mathcal E}(t)-\overline S_d(t)\le\log_2 C_n(L^{-1})
=2+O(L^{-1})$ uniformly in time,
with the same limiting ratio $d/L$.

\section{Monitored histories and entropy ordering}
\label{sec:up-prediction}
\subsection{A common output bound for the complete history}
\label{sec:SM_full_history}

Fix a finite nonadaptive history $\omega$ of Clifford gates and ordered
Pauli readouts, independent of the preparation and including cross-cut
gates. Each input follows its own legal Born outcomes.

\paragraph{Unsigned normal form.}
In the $2N$-dimensional binary symplectic Pauli space, a pure stabilizer
has Lagrangian labels $S=S^\perp$. Cliffords act symplectically; a Pauli
readout updates $S\mapsto\langle u\rangle+(S\cap u^\perp)$,
independently of its legal outcome. For each $\omega$ there exist a
symplectic $\gamma$ and isotropic $W$ that work for \emph{every} input $S$:

\begin{equation}
 T_\omega(S)=\gamma\bigl[W+(S\cap W^\perp)\bigr].
 \label{eq:SM_full_normal_form}
\end{equation}
Start with $(\gamma,W)=(I,0)$; a Clifford left-multiplies $\gamma$.
For a readout set $u=\gamma^{-1}P$ and $K=W+(S\cap W^\perp)$.
If $u\in W^\perp$, replace $W$ by $W+\langle u\rangle$, since
$\langle u\rangle+(K\cap u^\perp)
=W+\langle u\rangle+(S\cap W^\perp\cap u^\perp)$.
Otherwise choose $w\in W$ with $\langle u,w\rangle=1$.
The symplectic transvection $\tau_{u+w}(x)=x+\langle x,u+w\rangle(u+w)$
fixes $K\cap u^\perp$ and sends $w$ to $u$, because $K$ is isotropic.
The readout therefore sends $K$ to $\tau_{u+w}K$: replace
$\gamma$ by $\gamma\tau_{u+w}$, keeping $W$ fixed.
Both updates are input- and outcome-independent, completing the induction.

\paragraph{The intersection formula.}
Let the input sides have $n_A,n_B$ qubits and label spaces
$\mathcal A,\mathcal B$, with $N=n_A+n_B$. For any fixed output side
of $n_{\rm out}$ qubits and label space $\mathcal A_{\rm out}$, define
\begin{align}
 \begin{split}
 V_0&=\gamma^{-1}\mathcal A_{\rm out},\qquad F=(V_0+W)\cap W^\perp,\\
 c&=\dim(W\cap V_0),\qquad q(S)=\dim(S\cap F)-\dim(S\cap W).
 \end{split}\label{eq:SM_full_intersections}\\
 s_\omega(S)&=n_{\rm out}-c-q(S).
 \label{eq:SM_full_entropy}
\end{align}
Indeed, $W\subseteq F\subseteq W^\perp$; with $Q=S\cap W^\perp$,
quotienting $W+Q$ by $W$ gives
$\dim[(W+Q)\cap V_0]=c+\dim[Q\cap(V_0+W)]-\dim(Q\cap W)=c+q(S)$,
which proves the entropy formula.

\paragraph{A common bound for stabilizer inputs.}
Put $f=\dim F$, $f_A=\dim(F\cap\mathcal A)$ and
$f_B=\dim(F\cap\mathcal B)$. Every pure stabilizer input with entropy $e\le m$ obeys
$q(S)\ge b$, hence
\begin{align}
 b&=\max\{0,f-N,f_A-n_A-m,f_B-n_B-m\},
 \label{eq:SM_full_b}\\
 s_\omega(S)&\le M_\omega:=n_{\rm out}-c-b.
 \label{eq:SM_full_common_bound}
\end{align}
To prove $q(S)\ge b$, set
$r=\dim(S\cap W)$; $S=S^\perp$ gives
$\dim(S\cap W^\perp)=N-\dim W+r$. Intersecting with $F$ inside
$W^\perp$ gives $\dim(S\cap F)\ge f-N+r$, hence $q(S)\ge f-N$;
$W\subseteq F$ gives $q(S)\ge0$.

For the $A$ term, let $Q_0=S\cap W$ and $\pi_A$ project onto
$\mathcal A$. The pairing map $F\cap\mathcal A\to(\pi_AS)^*$
annihilates $\pi_AQ_0$, has kernel $S\cap F\cap\mathcal A$, and
$\dim\pi_AS=n_A+e$. Combining its kernel bound with
$Q_0+(S\cap F\cap\mathcal A)\subseteq S\cap F$ gives
\begin{align*}
 \dim(S\cap F\cap\mathcal A)&\ge f_A-n_A-e+\dim(\pi_AQ_0),\\
 q(S)&\ge f_A-n_A-e+\dim(\pi_AQ_0)-\dim(Q_0\cap\mathcal A)
 \ge f_A-n_A-m.
\end{align*}
The $B$ term follows identically.

\paragraph{Constant deficit of the uniform orbit.}
Let $S_{\rm orb}$ be the unsigned stabilizer space of a uniform
local Clifford orbit draw at entropy $m$.
For $n_A,n_B\ge1$, the probabilities that a specified nonzero
Pauli label lies in $S_{\rm orb}$ are
\begin{equation}
 \begin{split}
 \alpha_A&=\frac{2^{n_A-m}-1}{4^{n_A}-1},\qquad
 \alpha_B=\frac{2^{n_B-m}-1}{4^{n_B}-1},\\
 \zeta&=\frac{2^N-2^{n_A-m}-2^{n_B-m}+1}
 {(4^{n_A}-1)(4^{n_B}-1)}.
 \end{split}
 \label{eq:SM_full_orbit_probabilities}
\end{equation}
These are the $A$-only, $B$-only, and both-side probabilities:
transitivity in each class divides its stabilizer count (the numerator)
by its label count (the denominator). Thus
\begin{equation}
 \mathbb E|S_{\rm orb}\cap F|
 =1+\alpha_A(2^{f_A}-1)+\alpha_B(2^{f_B}-1)
 +\zeta(2^f-2^{f_A}-2^{f_B}+1).
 \label{eq:SM_full_count}
\end{equation}
For $m<\min(n_A,n_B)$, $\alpha_R\le2^{-n_R-m}$ and
$\zeta\le2^{-N}$. The latter is equivalent to
$2^{2n_A+n_B-m}+2^{n_A+2n_B-m}-2^N\ge4^{n_A}+4^{n_B}-1$,
which follows from $n_R-m\ge1$ and $4^{n_A}+4^{n_B}\ge2^{N+1}$.
Hence
\begin{equation}
 \mathbb E|S_{\rm orb}\cap F|
 \le1+2^{f_A-n_A-m}+2^{f_B-n_B-m}+2^{f-N}
 \le4\,2^b.
 \label{eq:SM_full_count_bound}
\end{equation}
At $n_A\ge n_B=m\ge1$, $\alpha_B=0$ and
$\zeta=2^{n_A-n_B}/(4^{n_A}-1)\le(4/3)2^{-N}$, so
Eq.~\eqref{eq:SM_full_count} is at most $(1+1+4/3)2^b<4\,2^b$;
interchange $A,B$ for the other endpoint. If one side is empty,
$m=0$ and its single nonzero label class gives $2\,2^b$ instead.
The same stronger bound holds at $m=n_A=n_B$, using
$\alpha_A=\alpha_B=0$, $\zeta=(2^N-1)^{-1}$, $b=[f-N]_+$.

The nonnegative deficit $\eta_{\rm orb}=M_\omega-s_{\rm orb}=q(S_{\rm orb})-b$,
where $s_{\rm orb}=s_\omega(S_{\rm orb})$,
therefore satisfies
\begin{align}
 \mathbb E\,2^{\eta_{\rm orb}}
 &=2^{-b}\mathbb E\frac{|S_{\rm orb}\cap F|}{|S_{\rm orb}\cap W|}
 \le2^{-b}\mathbb E|S_{\rm orb}\cap F|\le4,
 \label{eq:SM_full_moment}\\
 \mathbb E(s_{\mathcal E}-s_{\rm orb}\mid\omega)&\le2.
 \label{eq:SM_full_ordering}
\end{align}
The second line follows from Jensen, $\mathbb E\eta_{\rm orb}\le2$, and
Eq.~\eqref{eq:SM_full_common_bound} for any stabilizer reference ensemble
$\mathcal E$ with trajectory entropy $s_{\mathcal E}$ and initial
entropy at most $m$ on every realization. The average uses uniform encoding and each input's own legal Born
outcomes; unsigned evolution and entropy are independent of these signs.
Averaging over preparation-independent histories and reference preparations
gives $\overline S_{\mathcal E}(t)-\overline S_{\rm orb}(t)\le2$ at every
finite time, uniformly in history length, and hence
$\sup_t[\overline S_{\mathcal E}(t)-\overline S_{\rm orb}(t)]_+\le2$
for the mean curves.
\paragraph{Common fixed product ancillas.}
Suppose both inputs share the same fixed unsigned product-ancilla
space $C_0$, Lagrangian on the ancilla Pauli space. For variable-qubit
labels $S$, the normal form becomes
\begin{equation}
 T_\omega(C_0+S)
 =\gamma\bigl[C_0+W+(S\cap W^\perp)\bigr],
 \label{eq:SM_full_ancilla_normal}
\end{equation}
with $W,W^\perp$ in the variable space. If a pulled-back readout $u$
anticommutes with $C_0+W$, choose $w$ there with $\langle u,w\rangle=1$
and absorb $\tau_{u+w}$ into $\gamma$. Otherwise its ancilla component
lies in $C_0$; removing it leaves a variable label commuting with $W$,
which is adjoined to $W$. Thus the induction keeps $C_0$ fixed.

Let $V'$ be the image of $V_0\cap C_0^\perp$ in
$C_0^\perp/C_0$, identified with the variable space, where
$V_0=\gamma^{-1}\mathcal A_{\rm out}$. Quotienting by $C_0$, then $W$, gives
\begin{equation}
 \begin{split}
 s_\omega(S)&=C-q(S),\qquad F=(V'+W)\cap W^\perp,\\
 C&=n_{\rm out}-\dim(C_0\cap V_0)-\dim(W\cap V').
 \end{split}
 \label{eq:SM_full_ancilla_entropy}
\end{equation}
Here $q(S)=\dim(S\cap F)-\dim(S\cap W)$ as before;
only the input-independent constant changes. The proof requires
$W\subseteq F\subseteq W^\perp$, not nondegeneracy of $V'$, and applies
to unequal or empty variable sides with $M_\omega=C-b$.
Ancilla signs may differ: unsigned evolution and entropies are unchanged.

The common output bound and constant entropy deficit established here
are related to entanglement-of-assistance bounds discussed by
Hayden \emph{et al.}~\cite[Secs.~8--9]{Hayden2016Holographic}.
A related entropy-deficit structure occurs in random stabilizer tensor
networks~\cite[Apps.~A--B]{Nezami2020Multipartite}.
Here, direct orbit counting yields exact finite-size bounds for the
specified local Clifford encoding ensemble under complete nonadaptive
histories of Clifford gates and projective Pauli measurements, with each input's outcomes averaged
by its own Born probabilities.

\subsection{Two-design encoders and nonflat Schmidt spectra}
\label{sec:SM_twodesign_alltime}

The mean-ordering bound extends to arbitrary pure references and exact
unitary two-design encoders. Take two $n$-qubit halves of dimension
$Q=2^n$, $n\ge1$, and a pure seed $\ket{\psi}$ with Schmidt probabilities
$\lambda_j$. Write
$s=-\sum_j\lambda_j\log_2\lambda_j$ and
$s_2=-\log_2\sum_j\lambda_j^2$.

\begin{proposition}[Spectrum-sensitive all-time bound]
\label{prop:SM_twodesign_alltime}
Encode $\ket{\psi}$ by independent exact unitary two-designs on the two
halves and denote the resulting mean entropy by $\overline S_\psi(t)$.
For any preparation-independent, nonadaptive history of Clifford gates
and projective Pauli measurements, and any pure reference ensemble $\mathcal E$ of
mean initial entropy at most $s$, averaging each input with its own Born
probabilities gives
\begin{equation}
 \overline S_{\mathcal E}(t)-\overline S_\psi(t)
 \le 1+\log_2(1+2^{s-s_2})\le2+s-s_2
 \qquad\text{for all }t.
 \label{eq:SM_spectrum_alltime}
\end{equation}
\end{proposition}

Every flat spectrum has $s=s_2$, giving the two-ebit part of
Theorem~\ref{thm:ordering}; independent Haar encoders are included.

\paragraph{A common mean output bound.}
Fix a history, let $K_o$ denote its full-record Kraus operators, and set
\begin{equation}
 q_o=\frac{\Tr K_o^\dagger K_o}{Q^2},\qquad
 \ket{J_o}=\frac{\operatorname{vec}(K_o)}{\sqrt{\Tr K_o^\dagger K_o}}.
 \label{eq:SM_design_choi}
\end{equation}
Here $\operatorname{vec}$ uses the computational basis. Only nonzero
branches are retained. The normalized Choi state has
output systems $A_o,B_o$ and input systems $A_i,B_i$. It is a pure
stabilizer state, obtained by applying the history to halves of Bell
pairs. Unsigned updates do not depend on legal measurement signs, so
the following four entropies are independent of $o$:
\begin{align}
 e_0&=S_J(A_o),&e_3&=S_J(B_o),\notag\\
 e_A&=S_J(A_oA_i),&e_B&=S_J(A_oB_i).
 \label{eq:SM_design_cuts}
\end{align}
Define $M(x)=\min\{e_0,e_3,x+e_A,x+e_B\}$. We claim that every pure
input $\ket{\varphi}$ satisfies
\begin{equation}
 \sum_o p_\varphi(o)S_A(\ket{\varphi_o})\le M(S_A(\ket{\varphi})).
 \label{eq:SM_design_common_cap}
\end{equation}

Propagating any input Pauli operator $P$ through the fixed circuit gives
\begin{equation}
 K_oP=e^{i\theta(o,P)}Q_P K_{f_P(o)},
 \label{eq:SM_design_covariance}
\end{equation}
where $Q_P$ is an output Pauli operator, local across $A_o:B_o$, and $f_P$
permutes outcome strings.
Each measurement either preserves or flips its outcome label; the
fixed schedule makes this propagation independent of earlier outcomes.
Equation~\eqref{eq:SM_design_covariance} gives an exact teleportation
simulation of the flagged instrument \cite{Pirandola2017Fundamental}:
sample $\ket{J_o}$ with probability $q_o$ and Bell-teleport the two
actual input systems into the corresponding input ports. Including
the resource probability, the unnormalized output for teleportation
result $P$ is $Q^{-2}K_oP\ket{\varphi}$. Correct $Q_P$ and relabel
the record by $f_P$. Summing over the $Q^4$ teleportation results
recovers the original Born instrument.

For the partition $(A_oA_i):(B_oB_i)$, place actual input $A$ with
the first party and input $B$ with the second. This simulation uses
local operations and classical communication, so monotonicity of
mean pure-state entanglement bounds the output by
$S_A(\ket{\varphi})+e_A$. For $(A_oB_i):(B_oA_i)$, place actual input $B$
with the first party and $A$ with the second; the same argument gives
$S_A(\ket{\varphi})+e_B$. Also, each output marginal is supported on the
corresponding Choi marginal's support. The Choi marginal has a flat
spectrum of rank $2^{e_0}$ or $2^{e_3}$, yielding the remaining two bounds in
Eq.~\eqref{eq:SM_design_common_cap}. Since $M$ is increasing and
concave, a pure reference ensemble $\mathcal E$ of mean initial entropy at most
$s$ obeys $\overline S_{\mathcal E}\le M(s)$.

The cap agrees with $M_\omega$ in
Eq.~\eqref{eq:SM_full_common_bound}. In the notation of
Sec.~\ref{sec:SM_full_history}, the Choi stabilizers supported on
$A_o$, $A_oA_i$, and $A_oB_i$ have dimensions
$c$, $c+f_A$, and $c+f_B$, respectively. Together with
$f=2n_{\rm out}-c+\dim(W\cap V_0^\perp)$, this gives
\begin{align*}
 e_0&=n_{\rm out}-c,& e_3&=e_0-f+N,\\
 e_A&=e_0+n_A-f_A,& e_B&=e_0+n_B-f_B.
\end{align*}
Hence $M(m)=M_\omega$.

\paragraph{Born-normalized second moment.}
For one branch, let $\sigma_A(U)$ be its unnormalized output marginal,
$p(U)=\Tr\sigma_A(U)$, $q=\mathbb E p>0$, and
$P_2=\mathbb E\Tr\sigma_A(U)^2$, where the expectation is over encoders.
Then
\begin{equation}
 \mathbb E\bigl[pS(\sigma_A/p)\bigr]
 \ge q\log_2(q^2/P_2).
 \label{eq:SM_design_born_entropy}
\end{equation}
Indeed, write $d\nu=(p/q)d\mu$ for the posterior encoding law.
The R\'enyi-2 bound and Jensen give
\begin{align*}
 \mathbb E_\nu S(\sigma_A/p)
 &\ge-\mathbb E_\nu\log_2\frac{\Tr\sigma_A^2}{pq}
       +\mathbb E_\nu\log_2\frac pq\\
 &\ge-\log_2(P_2/q^2)+D(\nu\Vert\mu)
 \ge-\log_2(P_2/q^2).
\end{align*}
Zero-probability outcomes contribute zero.

Put $\tau=2^{-s_2}$. With $F_A,F_B$ the two-copy swaps, the exact local
twirl of $\ket{\psi_U}=(U_A\otimes U_B)\ket{\psi}$ is
$\mathbb E(\ket{\psi_U}\bra{\psi_U})^{\otimes2}
=a(I+F_AF_B)+b(F_A+F_B)$, where
\begin{align}
 A_\tau:=Q^4a&=\frac{Q^2(Q^2+1-2Q\tau)}{(Q^2-1)^2},\notag\\
 B_\tau:=Q^4b&=\frac{Q^2[(Q^2+1)\tau-2Q]}{(Q^2-1)^2}.
 \label{eq:SM_design_coefficients}
\end{align}
The swap trick identifies the four contractions with Choi purities:
\begin{equation}
 \frac{\mathbb E\Tr\sigma_{A,o}^2}{q_o^2}
 =A_\tau(2^{-e_0}+2^{-e_3})
  +B_\tau(2^{-e_A}+2^{-e_B}).
 \label{eq:SM_design_moment}
\end{equation}
Here $\mathbb E p_o=q_o$, since local two-design averaging gives the
maximally mixed input. If $B_\tau\ge0$, then $A_\tau\le1$ and
$B_\tau\le\tau$, so Eq.~\eqref{eq:SM_design_moment} is at most
$2^{-M(s)}(2+2^{s+1}\tau)$. If $B_\tau<0$, discard its negative
terms and use $A_\tau\le Q^2/(Q^2-1)\le4/3$. The resulting factor
$8/3$ is smaller than $2+2^{s+1}\tau\ge4$, since $s\ge s_2$.
Applying Eq.~\eqref{eq:SM_design_born_entropy} and summing
$\sum_o q_o=1$ therefore gives
\begin{equation}
 \overline S_\psi\ge M(s)-\log_2(2+2^{s+1}\tau).
 \label{eq:SM_design_high_entropy}
\end{equation}
Combining this with $\overline S_{\mathcal E}\le M(s)$ proves
Proposition~\ref{prop:SM_twodesign_alltime}. All bounds are independent
of history length; averaging preparation-independent random histories
preserves them.

\paragraph{Smoothing and independent nonflat pairs.}
Let a pure seed $\ket{\phi}$ on the same chain satisfy
$\frac12\|\,\ket{\psi}\bra{\psi}-\ket{\phi}\bra{\phi}\,\|_1
\le\epsilon$, with $0\le\epsilon\le1$. Write
$S_2(\ket{\phi})=-\log_2\Tr[(\Tr_B\ket{\phi}\bra{\phi})^2]$. Keep
$s=S_A(\ket{\psi})$ and the reference cap $M(s)$ fixed.
Repeating the second-moment argument for $\ket{\phi}$ bounds its deficit
from $M(s)$ by $2+[s-S_2(\ket{\phi})]_+$, since its moment coefficient is
at most $\max\{8/3,2+2^{s+1-S_2(\ket{\phi})}\}
\le4\max\{1,2^{s-S_2(\ket{\phi})}\}$.
Use the same encoders and retain both their labels and the complete
measurement record in a classical register $C$. Contractivity of
the recorded channel preserves trace distance at most $\epsilon$,
while mean output entanglement is $S(A|C)$. The classical-register
conditional-entropy continuity bound \cite[Lemma~2]{Winter2016Tight}
thus gives
\begin{equation}
 D_\psi(t)\le2+[s-S_2(\ket{\phi})]_++\epsilon n+g(\epsilon),
 \label{eq:SM_design_smooth}
\end{equation}
where $D_\psi=\overline S_{\mathcal E}-\overline S_\psi$,
$g(\epsilon)=(1+\epsilon)h_2(\epsilon/(1+\epsilon))$, and
$h_2(x)=-x\log_2x-(1-x)\log_2(1-x)$. The error is independent of
the number of records and the history length.

In particular, take $k$ copies of
$\sqrt\lambda\ket{00}+\sqrt{1-\lambda}\ket{11}$ and $n-k$
product pairs, with fixed $0<\lambda<1$, $k\to\infty$, and $n=O(k)$.
Any pure reference ensemble of mean initial entropy at most
$s=kh_2(\lambda)$ then satisfies
\begin{equation}
 \sup_t[D_\psi(t)]_+=O(\sqrt{k\ln k})=o(L),\qquad L=2n.
 \label{eq:SM_design_iid}
\end{equation}
For $\lambda\ne1/2$, set
$b_\lambda=|\log_2[\lambda/(1-\lambda)]|$ and
$u_k=b_\lambda\sqrt{2k\ln k}$. Discard Schmidt strings whose total
surprisal is below $s-u_k$. Hoeffding's inequality bounds their mass
by $\delta_k\le k^{-4}$. The normalized retained pure state $\ket{\phi_k}$
has trace distance $\sqrt{\delta_k}\le k^{-2}$ from the original seed
and every Schmidt probability is at most
$2^{-s+u_k}/(1-\delta_k)$; hence
$S_2(\ket{\phi_k})\ge s-u_k+\log_2(1-\delta_k)$.
For $k\ge2$, Eq.~\eqref{eq:SM_design_smooth} yields the uniform bound
\begin{equation}
 D_\psi(t)\le2+u_k-\log_2(1-k^{-4})
              +\frac n{k^2}+g(k^{-2}),
 \label{eq:SM_design_iid_explicit}
\end{equation}
which proves Eq.~\eqref{eq:SM_design_iid}. At $\lambda=1/2$, the
flat-spectrum two-ebit bound applies directly.

\subsection{Local comparison for the one-cycle lower bound}
\label{sec:SM_born_reduction}
\paragraph{The Born-weighted terminal law.}
Replace each cross-cut $U$ in the fixed history by the product Clifford
$V_A(U)\otimes V_B(U)$ chosen below. Each nonzero half-chain branch
has flat nonzero singular values~\cite[Theorem~4.2]{Kliuchnikov2023Verification}, hence
\begin{equation}
 K_{R,o}=c_{R,o}U_{R,o}\Pi_{z_R}V_{R,o},\qquad
 \Pi_z=(\ket{0}\bra{0})^{\otimes z}\otimes I,\qquad
 H_R=n-z_R=\log_2\operatorname{rank}K_{R,o},
 \label{eq:SM_history_normal_form}
\end{equation}
with local Cliffords $U,V$. From $I/2^n$, the sign-independent code
update $F_P(C)=(C\cap P^\perp)+\langle P\rangle$ gives the same $z_R$
on every nonzero branch, even for repeated or incompatible readouts;
$H_R$ is its conditional entropy.

For the signed uniform orbit,
$\mathbb E_\psi\ket{\psi}\bra{\psi}=I/2^{2n}$.
Put $\Pi=\Pi_{z_A}\otimes\Pi_{z_B}$, $\Lambda=z_A+z_B$, and write
$\ket{\psi_\Pi}$ for the normalized projection of $\ket{\psi}$.
Pauli relabelings permute the $2^\Lambda$ static outcomes and preserve
terminal entropy, so for every bounded $f$,
$\mathbb E_\psi\|\Pi\ket{\psi}\|^2 f(S_A(\ket{\psi_\Pi}))
=2^{-\Lambda}\mathbb E_{\rm static,Born}f(S_A)$.
Absorbing each branch's $V_{A,o}\otimes V_{B,o}$ into the orbit
and discarding its output local Cliffords yields the terminal law and mean
\begin{align}
 \mathbb E_\psi w_\psi(o)f(S_A(\ket{\psi_o}))
 &=|c_{A,o}|^2|c_{B,o}|^2 2^{-\Lambda}
   \mathbb E_{\rm static,Born}f(S_A),
 \label{eq:SM_terminal_law}\\
 \overline S^{\rm cut}_\omega
 &=\mathbb E_{\psi,\rm Born}S_A(\ket{\psi_o})
 =m-g_{n,m}(z_A,z_B).
 \label{eq:born-prediction-main}
\end{align}
Indeed, trace preservation makes the scalar factors sum to one ($f=1$);
summing for arbitrary $f$ gives the static terminal law.

\paragraph{The single-gate comparison cost.}
\label{sec:SM_boundary_cost}
The metric $d(S,T)=L-\dim(S\cap T)$ is invariant under common Cliffords
and nonincreasing under a common Pauli readout: for $R=S\cap T$, the
updated intersection contains $R$ if $P\in R$, and otherwise contains
$(R\cap P^\perp)\oplus\langle P\rangle$, of dimension at least $\dim R$.
Intersecting $R$ with the $A$ labels also gives
$|S_A(S)-S_A(T)|\le d(S,T)$.

Choose a product Clifford $V(U)$ minimizing
$q(U)=\operatorname{rank}_{\mathbb F_2}(\gamma_U-\gamma_{V(U)})$.
To prove $q\le2$, write the two-qubit symplectic matrix as
$\gamma=\left(\begin{smallmatrix}A&B\\C&D\end{smallmatrix}\right)$.
The blocks $B,C$ have equal rank: zero is local; at rank one,
$A,D$ are invertible and subtracting $\operatorname{diag}(A,D)$
leaves rank two. At rank two, multiply by
$\operatorname{diag}(B^{-1},C^{-1})$ to obtain
$\left(\begin{smallmatrix}A'&I\\I&\operatorname{adj}A'\end{smallmatrix}\right)$,
$\det A'=0$. Choose invertible $K$ with $A'+K$ invertible
($K^{-1}u\in\ker v^T$ for $A'=uv^T\ne0$; $K=I$ for $A'=0$).
Subtracting $\operatorname{diag}(A'+K,\operatorname{adj}(A'+K))$ leaves
\[
 \begin{pmatrix}K&I\\I&K^{-1}\end{pmatrix}
 =\begin{pmatrix}K\\I\end{pmatrix}
   \begin{pmatrix}I&K^{-1}\end{pmatrix},
\]
of rank two, since invertible binary $2\times2$ matrices are symplectic
and $\operatorname{adj}$ is linear in dimension two.

For a uniform two-qubit Clifford the distribution is
\begin{equation}
 \Pr(q=0)=\frac1{20},\qquad
 \Pr(q=1)=\frac9{20},\qquad
 \Pr(q=2)=\frac12,
 \qquad \mathbb E q=\frac{29}{20}.
 \label{eq:boundary-gate-cost}
\end{equation}
Indeed, $|\operatorname{Sp}(4,2)|=720$ and its local subgroup has 36
elements. Rank-one symplectic perturbations are transvections
$T_v:w\mapsto w+\langle v,w\rangle v$. The nine both-side Paulis give
distinct local cosets: for $v=(a,b)$, $w=(c,d)$, the off-diagonal block
of $T_wT_v$ is $[(a+\langle w,v\rangle c)b^T+cd^T]J$, zero only if
$a=c,b=d$. Thus $(q=0,1,2)$ has $(36,324,360)$ elements, regardless of signs.

At a substituted gate,
$d(\gamma_US,\gamma_VT)\le q(U)+d(S,T)$; later common operations
are nonexpansive. For histories differing only at this gate, every stabilizer input,
with its own legal Born signs, obeys
\begin{equation}
 |s^{\rm full}(t)-s^{\rm cut}(t)|\le q(U)\le2.
 \label{eq:up-boundary-sm}
\end{equation}
The mean single-gate cost is $29/20$ per input.

\subsection{First-readout and one-cycle reversal}
\label{sec:SM_analytic_switch}
For even $n,m$, compare $\ket{\Phi_m^0}$ and
$\ket{\Phi_m^{\rm orb}}\sim\Omega(n,n,m)$ against the same reference
$\ket{\Phi_{m'}^{\rm orb}}\sim\Omega(n,n,m')$, $m'<m$, under the
main-text protocol. Write $s=m/L$, $s'=m'/L$, and let $f(p)=2G_0(p)/m$
denote the polynomial in Eq.~\eqref{eq:d0-quartic}.

\paragraph{First-readout ordering.}\label{sec:ts-switch}
Take $L\in64\mathbb N$, $p=1/5$, $m=3L/8$, and $m'=21L/64$;
also consider the two-design preparation $\ket{\Phi_m^{\rm des}}$.
Any fixed first layer preserves the encoders' Haar two-copy moments
and the orbit laws. Proposition~\ref{prop:ts-bilateral} and
Sec.~\ref{sec:ts-window} therefore give exponentially vanishing encoded
losses, since both redundancy fractions $1/4,11/32$ exceed $p$.
With $f(1/5)=2218/3125$,
\begin{align}
 D_d(1/2)&=G_d-G_{\rm ref}-(m-m'),
 \label{eq:ts-Q}\\
 \frac{D_0(1/2)}L&\longrightarrow0.086205,\qquad
 \frac{D_{\rm des}(1/2)}L\longrightarrow-3/64.
 \label{eq:ts-switch}
\end{align}
Here $G_0/L=3327/25000=0.13308$.
At $L=256$, the losses $(G_{\rm ref},G_{\rm orb},G_0)$ are
approximately $(0.00115,0.524,34.07)$ ebits, giving the line
$(G_{\rm ref}+12)/256$ in Fig.~\ref{fig:full-response}(a).

For the same main parameters and $d/L\to r<15/64$,
Eqs.~\eqref{eq:depth-random-front} and \eqref{eq:ts-Q} give
\[
 \liminf_{L\to\infty}\frac{D_d(1/2)}L
 \ge\frac15\left(\frac38-\frac{3r}{5}\right)-\frac3{64}
 =\frac9{320}-\frac{3r}{25}>0,
\]
since the reference loss vanishes.
At fixed entropy densities and $p\in[0,1]$,
Eqs.~\eqref{eq:up-threshold} and \eqref{eq:d0-main} give
\[
 \frac{D_0(1/2)}L\longrightarrow
 \min\{s',(1-p)/2\}-s[1-f(p)/2].
\]
For the main densities, this limit is positive between the root of
$(1-p)^2(1+p^2/5)=7/8$, approximately $0.06498$, and $p=1$.

\paragraph{One-cycle reversal.}
The first readout leaves mean entropy $m-G_0(p)$. The crossing gate's
local substitute preserves entropy, so Eq.~\eqref{eq:boundary-gate-cost}
gives $\overline S_0^{\rm full}(1)\le m-G_0(p)+29/20$.
For the contiguous seed with $n-m\ge2$, each adjacent first-layer
$\ket{00}$ pair independently retains a Bell pair with probability
$q=(2/5)(1-p)^2$. If zero, one, or both cut endpoints remain entangled with their neighbors
in the same half, the crossing gate adds $2/5$, $4/5$, or $29/20$ ebits
on average, respectively.
The first counts 6 Bell states among 15 unsigned two-qubit stabilizers;
for one purifier, only 3 of the 15 images of its sole check leave the
opposite endpoint pure. With two purifiers, entropy is the Choi state's
cross-block rank. Rank one occurs for $9\cdot6^2=324$ matrices
(choose rank-one $B$ and invertible $A,D$, which fix $C$); 36 local
matrices have rank zero and 360 rank two, giving $29/20$. Averaging gives
\begin{equation}
 c(p)=\frac25+\frac8{25}(1-p)^2+\frac1{25}(1-p)^4,
 \qquad
 \overline S_0^{\rm full}(1)
 \le m-G_0(p)+c(p).
 \label{eq:switch-high-one-cycle}
\end{equation}
Other gates in that layer are local and commute with the crossing gate;
the final readout cannot increase mean entropy. At $m=n$, replace $c(p)$ by $29/20$.

Each half has $L$ readout opportunities, with independent measurement counts
$N_A,N_B\sim\operatorname{Bin}(L,p)$. Each readout adds at most one
input check in the local protocol. Thus for
$V=\min\{m',(n-\max(N_A,N_B))_+\}$,
\begin{equation}
 \overline S_{\rm ref}^{\rm full}(1)
 \ge\mathbb EV-\frac32-\frac{29}{20}
 =\mathbb EV-\frac{59}{20}.
 \label{eq:switch-event-capacity}
\end{equation}
With $F_{L,p}$ the binomial cumulative distribution,
\begin{equation}
 \mathbb EV=\sum_{r=1}^{m'}F_{L,p}(n-r)^2,\qquad
 D_0(1)\ge\mathbb EV-m+G_0(p)-\frac{22}{5}.
 \label{eq:one-cycle-exact-binomial}
\end{equation}
For $n-m\ge2$, replace $22/5$ by $59/20+c(p)$.
The rank-bound function is 1-Lipschitz and
$\mathbb E(\max_RN_R-Lp)_+\le\sqrt{Lp(1-p)}$, so
$\mathbb EV\ge\min\{m',(n-Lp)_+\}-\sqrt{Lp(1-p)}$. Hence
\begin{equation}
 \begin{gathered}
 D_0(1)\ge L\Gamma-\sqrt{Lp(1-p)}-\frac{22}{5},\\
 \Gamma=\min\{s',(1/2-p)_+\}-s[1-f(p)/2].
 \end{gathered}
 \label{eq:switch-gamma}
\end{equation}
Again $22/5$ sharpens to $59/20+c(p)$ for $n-m\ge2$.
Thus $\Gamma>0$ gives extensive reversal; bounded rounding
$m=sL+\epsilon$, $m'=s'L+\epsilon'$ costs at most
$|\epsilon'|+[1-f(p)/2]|\epsilon|$.
At $s=3/8$ and $s'=21/64$, $\Gamma$ increases up to $p=11/64$
and then decreases on $[11/64,1/4]$. It is positive at both
$p=0.065$ and $p=1/4$.
At $p=1/5$, $m=3L/8$, $m'=21L/64$, $L\in64\mathbb N$,
$c(1/5)=9706/15625$ gives
\begin{equation}
 D_0(1)\ge\frac{363L}{6250}-\frac25\sqrt L
                    -\frac{223199}{62500}.
 \label{eq:switch-rational}
\end{equation}
Since $223199/62500<18/5$, this implies the main-text bound.
This is positive for allowed $L\ge192$, whereas
Eq.~\eqref{eq:one-cycle-exact-binomial} suffices at $L=128$;
the limiting lower density is $0.05808$.

\paragraph{Two-design references.}
The lower reference ensemble $\mathcal E$ may instead consist of draws
$\ket{\Phi_{m'}^{\rm des}}$, with Haar encoders among the choices.
Apply the single-gate comparison to the normalized Choi states of
the full and local-substitute histories. Each entropy in
Eq.~\eqref{eq:SM_design_cuts} changes by at most $q(U)$, so their
caps $M(m')$ differ by at most $q(U)$.
For the local history, Eq.~\eqref{eq:SM_history_normal_form} gives
$M(m')=\min\{m',H_A,H_B\}\ge V$.
Equation~\eqref{eq:SM_design_high_entropy} therefore yields
\begin{equation}
 \overline S_{\mathcal E}^{\rm full}(1)
 \ge\mathbb EV-2-\mathbb E q
 =\mathbb EV-\frac{69}{20}.
 \label{eq:switch-design-reference}
\end{equation}
With $\ket{\Phi_{m'}^{\rm des}}$ as reference, the lower bound in
Eq.~\eqref{eq:switch-rational} decreases by $1/2$ ebit, leaving its
extensive coefficient unchanged.

\paragraph{Typical trajectory separation.}
\label{sec:SM_typical}
For stabilizer preparations with reference ensemble $\mathcal E$ of
initial entropy at most $m$ on every realization,
independent of shared gates and masks $\omega$, and each input's own
legal Born record, the deficit in trajectory entropy
$\eta_t=M_\omega-s_d(t)\ge s_{\mathcal E}(t)-s_d(t)$ satisfies
$\mathbb E(2^{\eta_t}\mid\omega)\le C$: $C=4$ for a uniform orbit
[Eq.~\eqref{eq:SM_full_moment}], or $C=C_n(x)<16$ for local encoders
at the sufficient depth [Eqs.~\eqref{eq:depth-alltime-moment} and
\eqref{eq:depth-linear-family}]. Markov and a union bound give, for
any prescribed finite time set $\mathcal T$, 
\begin{equation}
 \Pr\!\left\{\max_{t\in\mathcal T}[s_{\mathcal E}(t)-s_d(t)]
                  \ge\epsilon L\right\}
 \le C|\mathcal T|2^{-\epsilon L}.
 \label{eq:trajectory-protection}
\end{equation}
This is exponentially small for fixed $\epsilon>0$ and
$\log|\mathcal T|=o(L)$, and includes encoding randomness;
the all-time mean bound instead takes the supremum after averaging.

For $\ket{\Phi_m^0}$, the first loss $X$ sums $m/2$ independent block losses
in $[0,2]$, with $\mathbb EX=G_0(p)$. Let $Z=\max_R(N_R-Lp)_+$
and $\eta_{\rm ref}$ be the lower orbit's one-cycle deficit in the local protocol.
Restoring the crossing gate costs the reference at most two ebits.
The second layer adds at most two ebits to the unencoded trajectory,
and local readouts cannot increase branch entropy. The rank estimate
of Eq.~\eqref{eq:switch-gamma} gives
\[
 s_{\rm ref}(1)-s_0(1)
 \ge L\Gamma-Z-(G_0(p)-X)-\eta_{\rm ref}-4.
\]
For $0<\epsilon<\Gamma$, $a=(\Gamma-\epsilon)/4$ and $aL\ge4$,
a union bound using the Chernoff bound for the counts, Hoeffding's
inequality for $X$, and $\mathbb E2^{\eta_{\rm ref}}\le3$ gives
\begin{equation}
 \Pr\{s_{\rm ref}(1)-s_0(1)<\epsilon L\}
 \le2e^{-L\operatorname{kl}(p+a\Vert p)}
       +e^{-a^2L^2/m}+3\,2^{-aL}.
 \label{eq:trajectory-reversal}
\end{equation}
At fixed densities
with $\Gamma>0$, unencoded one-cycle reversal and encoded protection
for $\log|\mathcal T|=o(L)$ therefore hold jointly with probability
$1-e^{-\Omega(L)}$, for a common lower orbit and shared history.
These concern trajectory ordering; the stationary-distance QME
certificate uses ensemble means.

\subsection{Ordering after the first readout of the unencoded state}
\label{sec:SM_no_return}
For $L\in64\mathbb N$, $m=3L/8$, $m'=21L/64$, and $p=1/5$,
take the reference ensemble $\mathcal E$ of draws $\ket{\Phi_{m'}^{\rm des}}$,
independently of the monitoring history. Let $X$ be the first-readout loss of
$\ket{\Phi_m^0}$ and $\Delta=m-m'=3L/64$.
By Eq.~\eqref{eq:d0-quartic}, $X$ sums $m/2$ independent $[0,2]$
block losses with $\mathbb EX=(3327/25000)L$, so Hoeffding gives
\begin{equation}
 \Pr\{X<\Delta\}
 \le \exp\!\left[-\frac{(0.086205L)^2}{m}\right].
 \label{eq:SM_no_return_hoeffding}
\end{equation}

Fix the gates and masks through time $t\ge1/2$.
The stabilizer loss $X$ is independent of the first outcome signs.
On $X\ge\Delta$, every first-readout branch has entropy
$m-X\le m'$ and is an eigenstate of the measured $Z$ operators.
Undoing the first local gate layer gives an input of the same
entropy whose first readout produces that branch with certainty.
Its subsequent evolution is therefore that of the corresponding
branch from $\ket{\Phi_m^0}$.
Equation~\eqref{eq:SM_design_common_cap}, applied to the complete
history, bounds each branch's subsequent Born-averaged entropy by
$M(m')$. Averaging the first outcomes preserves this bound.
Meanwhile, Eq.~\eqref{eq:SM_design_high_entropy} gives
$\overline S_{\mathcal E}(t)\ge M(m')-2$ for this fixed history.

Thus the conditional mean difference is at least $-2$ on
$X\ge\Delta$. On the remaining histories it is bounded below by
$-L/2$, yielding
\begin{equation}
 \overline S_{\mathcal E}(t)-\overline S_0(t)\ge-2-\frac L2
 \exp\!\left[-\frac{(0.086205L)^2}{m}\right]
 =-2-O(Le^{-cL}),\qquad t\ge\tfrac12,
 \label{eq:SM_no_return}
\end{equation}
for some $c>0$.
The time-uniform constant excludes a later negative difference of
fixed extensive amplitude after the one-cycle reversal.

\section{Stationary entropy and QME certificates}
\label{sec:up-stationary}
\subsection{An all-input stationary bound}
\label{sec:SM_stationary_envelope}
Evolve the maximally mixed stabilizer code $M_0=\{0\}$ ($I/2^L$).
Clifford conjugation and the full commuting-subspace update
$F_P(C)=(C\cap P^\perp)+\langle P\rangle$ preserve inclusion, including
products of rows anticommuting with $P$. Hence $M\subseteq S$ for every pure
stabilizer input along its own legal Born record. For an independent
terminal generator basis $G$ of rank $r$, with restrictions $G_A,G_B$,
\begin{equation}
 h_A=n-r+\operatorname{rank}G_B,\quad
 h_B=n-r+\operatorname{rank}G_A,\quad
 S_A\le u:=\min(h_A,h_B)\in[0,n].
 \label{eq:SM_stationary_envelope}
\end{equation}
For any pure input and any record $o$ of nonzero probability,
$K_o\ket{\psi}$ lies in the support of $K_oK_o^\dagger$, the output
from the maximally mixed state up to normalization. The local checks of that mixed
state also stabilize the pure output. Since
$\dim M_A=r-\operatorname{rank}G_B$, these checks give $S_A\le h_A$.
Purity gives the corresponding bound from $B$.
For finite $L$ and $0<p<1$, full Clifford support and finite no-readout
sequences connect all pure stabilizer states; identity gates with no
readouts give a self-loop. The integer-cycle chain thus has a unique
stationary law and transported half-cycle phases.
Every pure input enters this chain almost surely: a full $Z$ readout
occurs with probability $p^L$ per layer and produces a
computational-basis state. A stationary input
independent of the next $T$ cycles retains its law, proving
\begin{equation}
 S_\infty\le\mathbb E u(T),
 \label{eq:SM_stationary_phase}
\end{equation}
where $S_\infty$ is the post-even-bond-readout mean.

Samples have theoretical supports and fixed sizes and durations, retaining
histories that purify early. For independent $X_i\in[0,1]$ of common mean
$\mu$, convexity gives
$\mathbb E e^{\lambda X_i}\le1-\mu+\mu e^\lambda$
\cite[Lemma~9]{GarivierCappe2011KL}, hence
$\Pr(\bar X\le x)\le e^{-N\operatorname{kl}(x\Vert\mu)}$ for $x<\mu$
and the analogous upper-tail bound. Here
$\operatorname{kl}(x\Vert q)=x\ln(x/q)+(1-x)\ln[(1-x)/(1-q)]$.
For the interior means below, the roots $q_-\le x\le q_+$ of
$N\operatorname{kl}(x\Vert q)=\ln(1/\alpha)$ are one-sided
$1-\alpha$ confidence bounds; multiplication by $b$ covers $[0,b]$.
For $Z\in[a,b]$, the Hoeffding lower bound is
$\bar Z-(b-a)\sqrt{\ln(1/\alpha)/(2N)}$, truncated to $[a,b]$
\cite{Hoeffding1963Bounds}.

\subsection{QME certificates for the pair of Fig.~\ref{fig:phenomenon-mechanism}}
\label{sec:SM_stationary_L256}
At $p=1/5$, let $h_1,e_1$ be the full-circuit means at $t=1$ for
$\ket{\Phi_{3L/8}^0}$ and $\ket{\Phi_{21L/64}^{\rm orb}}$, respectively.
Equation~\eqref{eq:switch-event-capacity} and the binomial estimate
preceding Eq.~\eqref{eq:switch-gamma} give
\begin{equation}
 \frac{h_1+e_1}{2}\ge\frac{e_1}{2}
 \ge\frac{3L}{20}-\frac15\sqrt L-\frac{59}{40}.
 \label{eq:SM_asymptotic_midpoint}
\end{equation}
Together with Eq.~\eqref{eq:switch-on-main}, this proves one-cycle QME
for all sufficiently large multiples of 64 if $S_\infty(L)=o(L)$,
which also puts the stationary mean below $21L/64$. Only the stationary
scaling is conditional; both transient bounds are analytic.
More generally, $\limsup_{L\to\infty}S_\infty(L)/L<3/20$
suffices for one-cycle QME.
The contiguous-seed refinement of Eq.~\eqref{eq:one-cycle-exact-binomial}
gives the following bounds in ebits. Upper bounds are rounded up and lower
bounds down. The midpoint bound uses $h_1\ge0$.
\begin{equation}
 \begin{array}{c|rrrr}
 L&h_1\ \text{upper}&e_1\ \text{lower}
   &e_1-h_1\ \text{lower}&(e_1+h_1)/2\ \text{lower}\\\hline
 128&31.587&32.840&1.253&16.420\\
 256&62.553&70.215&7.662&35.107
 \end{array}
 \label{eq:s13-midpoint-256}
\end{equation}
For $L=128,256$, separate samples of 128 independent full gate/mask
histories at $p=0.20$ have $(T,\sum_i u_i)=(32,194)$ and $(64,159)$,
respectively, with envelope support $[0,L/2]$. We use $\alpha=0.005$
at both sizes.
For outcomes in $[0,b]$ with sample mean $\bar u$, the one-sided KL
upper limit is $bq_+(\bar u/b;N,\alpha)$. With $b=L/2$ and $N=128$,
the observed upper limits in ebits are
\begin{align}
 U_{128}^{\rm obs}&=64q_+\!\left(\frac{194}{128\cdot64};128,0.005\right)<6.10,
 \label{eq:s13-upper}\\
 U_{256}^{\rm obs}&=128q_+\!\left(\frac{159}{128^2};128,0.005\right)<8.74.
 \label{eq:s13-upper-256}
\end{align}
Both certificates lie below the initial lower entropy and the one-cycle
midpoint; since $e_1>h_1$,
$(e_1-S_\infty)^2-(h_1-S_\infty)^2
=(e_1-h_1)(e_1+h_1-2S_\infty)>0$.
The initial distance order is therefore reversed at $t=1$
[Eq.~\eqref{eq:strict-qme}] at each size with one-sided confidence
of at least $99.5\%$. The two certificates hold jointly with confidence
at least $99\%$ by the union bound.
The same confidence events give $e_1>S_\infty$, so
\[
 |e_1-S_\infty|-|h_1-S_\infty|
 =\min\{e_1-h_1,e_1+h_1-2S_\infty\}.
\]
The bounds above certify mean-distance advantages of at least
$1.253$ and $7.662$ ebits for $L=128$ and $256$, respectively.
For a two-design reference, Eq.~\eqref{eq:switch-design-reference}
decreases the one-cycle lower bounds on $e_1$ and $e_1-h_1$ by
$1/2$ ebit, and the midpoint bounds by $1/4$ ebit; both finite-size
QME certificates remain valid on the same confidence events.

\subsection{Entropy-distribution distances}
\label{sec:SM_entropy_distribution}
Let $\mu_d(t)$ and $\mu_{\rm ref}(t)$ be the trajectory-entropy laws
at integer-cycle $t$, and $\mu_\infty$ the common stationary law.
For Wasserstein-1 cost $|x-y|$,
nonnegativity gives $W_1(\mu_d(t),\delta_0)=\overline S_d(t)$ and
$W_1(\mu_\infty,\delta_0)=S_\infty$. The triangle inequality implies
$|W_1(\mu_d(t),\mu_\infty)-\overline S_d(t)|\le S_\infty$, hence
\begin{equation}
 \left|W_1(\mu_{\rm ref}(t),\mu_\infty)-W_1(\mu_d(t),\mu_\infty)-D_d(t)\right|
 \le 2S_\infty.
 \label{eq:SM_wasserstein_mean}
\end{equation}
For the analytic pair of Theorem~\ref{thm:ordering}, the initial laws
are point masses at $m>m'$, with the same distance ordering when
$m-m'>2S_\infty$. The unencoded one-cycle distance difference is at
least $363L/6250-(2/5)\sqrt L-18/5-2S_\infty$.
For $\ket{\Phi_m^{\rm des}}$ in Theorem~\ref{thm:ordering},
including Haar encoding, and any pure reference ensemble $\mathcal E$ with
mean initial entropy at most $m$ and trajectory-entropy law $\mu_{\mathcal E}(t)$,
\begin{equation}
 W_1(\mu_{\mathcal E}(t),\mu_\infty)-W_1(\mu_{\rm des}(t),\mu_\infty)
 \le2+2S_\infty
 \label{eq:SM_wasserstein_alltime}
\end{equation}
at every integer-cycle time; sufficient local depth replaces $2$ by
$\log_2C_n(x)<4$.
The same upper bounds hold for absolute mean deviations because
$\bigl||z-S_\infty|-z\bigr|\le S_\infty$ for $z\ge0$.
For the analytic pair, if $S_\infty=o(L)$, the initial ordering holds and the unencoded state
reaches every fixed reversal density $0<\epsilon<363/6250$ at one cycle
for sufficiently large allowed $L$. Exact two-design and sufficient-depth local
encoding exclude fixed positive reversal density at every time in both
measures. The finite-size certificates above concern the mean-entropy
criterion~\eqref{eq:strict-qme}.

\section{Numerical methods and controls}
\label{sec:SM_methods}
\subsection{Circuit protocol and sampling}
\label{sec:SM_circuit}
Stim tableaux \cite{Aaronson2004Improved,Gidney2021Stim} use independent
uniform gates among the 11520 two-qubit Cliffords modulo global phase.
Times $t=0,0.5,1,\ldots$ denote initial or post-readout states.
Restricted-generator ranks give $S_A$ independently of legal Born signs
\cite{Fattal2004Entanglement}; the single-site sum $\chi$ is evaluated as in
Sec.~\ref{sec:SM_analytical}. Paired resampling retains every condition
and time in each independent trajectory or preparation block, after
averaging replicas within blocks.

\paragraph{Original-preparation illustration.}
Figure~\ref{fig:phenomenon-mechanism}(a) uses the original $L=128$,
$p=0.20$ sample: 1000 trajectories each for complete rainbow,
unconditional depth-$L$ Clifford, three-regular graph and product
preparations, with trajectory standard errors. These differ from the
fixed-spectrum preparations in panels (b--d).

\paragraph{Short-time paired trajectories.}
For each $L=64,128,256,512,1024$, 1000 independent units pair
$\ket{\Phi_m^0}$, $\ket{\Phi_m^{\rm orb}}$, and
$\ket{\Phi_{m'}^{\rm orb}}$ at $p=0.20$ on $t=0,0.5,\ldots,8$.
Here $m=3L/8$ and $m'=21L/64$.
$L=256$ also includes the product state. Each unit has independent
orbit preparations and monitoring history. Standard errors and 2000
paired trajectory resamples retain within-unit correlations. This sample
is separate from the preparation-block design below. Across all sizes,
the two orbit means differ by at most $0.031$ ebit on $t=1.5,2,\ldots,8$,
and the orbit's largest positive mean difference on the full short grid
is at most $0.025$ ebit.

\paragraph{Fixed-spectrum preparation and sampling.}
\label{sec:SM_encoding}
For $L=64,128,256$, independent within-half Clifford circuits encode
$\ket{\Phi_m^0}$ with $m=3L/8$, using nested prefixes
$r=d/L=0,1/8,1/4,1/2$ in half-layers. Independent exact uniform
symplectic tableaux give $\ket{\Phi_m^{\rm orb}}$ and the reference
$\ket{\Phi_{m'}^{\rm orb}}$ with $m'=21L/64$. Preparation and
monitoring seeds are independent. Each size has 100 independent blocks, each containing
one reference, five high-entropy preparations and 20 monitoring replicas
at $p=0.20$. Within a replica, conditions share gate and mask streams
and the uniform random numbers used to sample outcomes, with each
input using its own conditional Born probabilities.

In each of ten thousand paired block bootstrap resamples, we reconstruct
$D_d(t)=\overline S_{\rm ref}(t)-\overline S_d(t)$ and recompute the peak
$\Delta_d=\max_{t=0,0.5,\ldots,256}D_d(t)$ for $d=0,L/8,L/4,L/2$
and the orbit.
The 95th percentile of
$\max_t|\widehat D_{d,b}^*(t)-\widehat D_d(t)|$ sets each preparation's
nominal 95\% simultaneous band. A crossing is resolved if the lower
edge exceeds zero at any sampled time.
Figure~\ref{fig:phenomenon-mechanism}(c) uses pointwise percentile
intervals and (d) standard errors of these bootstrap peak estimates. Since the first gates
are intra-half, averaging preparations, gates, masks and Born outcomes gives
\begin{equation}
 D_d(1/2)=G_d(p)-G_{\rm ref}(p)-3L/64.
 \label{eq:SM_first_readout_margin}
\end{equation}
Table~\ref{tab:SM_shell_crossing} uses the measurement masks from the
full trajectories. It reports pointwise first-readout intervals for each
preparation. The crossing test uses the simultaneous time band.

\setcounter{table}{0}
\begin{table}[!htbp]
\caption{Uniform local Clifford reference. The upper line gives $\Delta_d/L$. Parentheses give the paired block-bootstrap standard error in units of the last digit shown. The lower line gives $D_d(1/2)$ in ebits with its pointwise 95\% interval. The peak is a maximum over 513 sampled times and can be biased upward.
Crossings are assessed with the simultaneous time band.}
\label{tab:SM_shell_crossing}
\centering\begin{ruledtabular}\begin{tabular}{cccc}
Preparation & $L=64$ & $L=128$ & $L=256$\\
\raisebox{5pt}{$\ket{\Phi_m^0}$} & \shortstack{$0.1177(6)$\\$5.20\ [5.11,5.29]$} & \shortstack{$0.1319(5)$\\$10.96\ [10.82,11.10]$} & \shortstack{$0.1408(3)$\\$22.03\ [21.80,22.26]$}\\[2pt]
\raisebox{5pt}{$\ket{\Phi_m^{L/8}}$} & \shortstack{$0.0464(6)$\\$2.34\ [2.26,2.42]$} & \shortstack{$0.0482(4)$\\$3.84\ [3.74,3.95]$} & \shortstack{$0.0517(3)$\\$6.04\ [5.87,6.20]$}\\[2pt]
\raisebox{5pt}{$\ket{\Phi_m^{L/4}}$} & \shortstack{$0.0237(6)$\\$1.12\ [1.05,1.19]$} & \shortstack{$0.0249(4)$\\$1.67\ [1.56,1.77]$} & \shortstack{$0.0284(3)$\\$2.25\ [2.09,2.39]$}\\[2pt]
\raisebox{5pt}{$\ket{\Phi_m^{L/2}}$} & \shortstack{$0.0022(3)$\\$-0.72\ [-0.80,-0.64]$} & \shortstack{$0.00079(20)$\\$-2.16\ [-2.26,-2.05]$} & \shortstack{$0.00013(4)$\\$-5.38\ [-5.52,-5.23]$}\\[2pt]
\raisebox{5pt}{$\ket{\Phi_m^{\rm orb}}$} & \shortstack{$0.00029(13)$\\$-2.12\ [-2.16,-2.07]$} & \shortstack{$0.00018(7)$\\$-5.14\ [-5.20,-5.08]$} & \shortstack{$0.00012(3)$\\$-11.46\ [-11.52,-11.39]$}\\
\end{tabular}\end{ruledtabular}
\end{table}

The same resamples give joint one-sided peak limits for
$\ket{\Phi_m^0}$, $\ket{\Phi_m^{L/2}}$, and $\ket{\Phi_m^{\rm orb}}$
from the 95th percentile of
$\max\{\max_t(\widehat D_{0,b}^*-\widehat D_0),
\max_t(\widehat D_{L/2}-\widehat D_{L/2,b}^*),
\max_t(\widehat D_{\rm orb}-\widehat D_{{\rm orb},b}^*)\}$.
At $L=256$, the joint limits put the peak at or above $35.82$ ebits for
the unencoded preparation and at or below $0.240$ ebit for both
depth $L/2$ and the orbit.
These limits give at least 149-fold suppression on the sampled grid at
nominal 95\% joint confidence.
The joint limits compare peak heights. The per-preparation bands test for crossings.

\paragraph{Complete first response.}\label{sec:SM_full_record_production}
At $L=256$, 100 high-state blocks with 20 first-gate replicas each use
32 fresh rows of 256 independent uniform random numbers on the unit
interval per replica to generate masks independently of those in
Table~\ref{tab:SM_shell_crossing}.
The masks $J(p)=\{i:u_i<p\}$ are shared across preparations and nested across
$p=0,0.01,\ldots,0.20$. For block draws $\ket{\Phi_{m,b}^d}$, record ranks give
$\widehat G_d(p)=100^{-1}\sum_b(20\cdot32)^{-1}
\sum_{g,k}\kappa_{J_{bgk}(p)}(U_{0,bg}\ket{\Phi_{m,b}^d})$.
Ten thousand paired block resamples give pointwise percentile intervals
with linear interpolation and no additional conditional-mask standard
error; the linear term averages exact post-gate one-site memberships.
Analytic endpoints use 401 $p$ points, numerical responses 21.
Figures~\ref{fig:SM_pilots}(a) and \ref{fig:full-response}(a) share the
exact-recursion threshold $[G_{\rm ref}(p)+12]/256$.

\subsection{Transition location and first-readout rate dependence}
\label{sec:SM_pilots}
Starting from $\ket{+}^{\otimes L}$, we average the
tripartite mutual information of contiguous quarters
\cite{Zabalo2020Critical,Gullans2020Dynamical,Li2019Measurement},
$I_3(Q_1{:}Q_2{:}Q_3)=\sum_iS_{Q_i}-\sum_{i<j}S_{Q_iQ_j}
+S_{Q_1Q_2Q_3}$, over each trajectory's last recorded quarter
(cycles $3L+1$--$4L$ near criticality for $L\ge64$). Each cell has
500 trajectories at $L=32,64,128,256,512$ and
$p=0.14,0.15,0.16,0.17,0.18$. Bootstrap indices are shared across $p$
within each size and independent across sizes. Crossings for adjacent
sizes $(32,64),(64,128),(128,256),(256,512)$ are respectively
$0.1565\pm0.0012$, $0.1592\pm0.0006$, $0.1600\pm0.0013$, and
$0.1614\pm0.0024$ (bootstrap standard errors). Their mean is $0.1593$
with bootstrap standard error $0.0006$ and 95\% interval
$[0.1586,0.1608]$. The uncertainty in $p_c=0.159(3)$ also includes
finite-size drift.

\begin{figure}[!htbp]
\centering\includegraphics{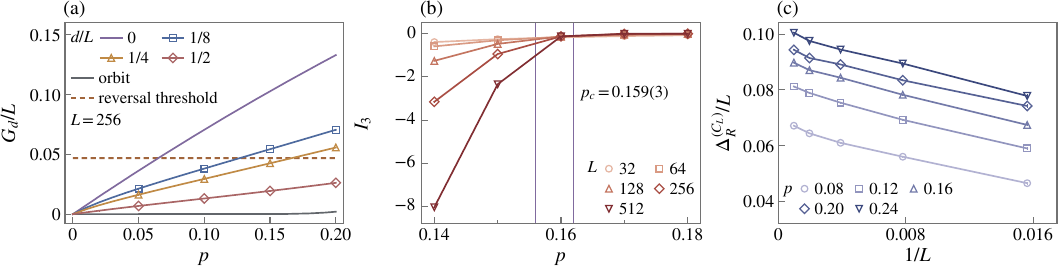}
\caption{First-readout response and measurement-rate checks.
(a) Complete-record loss $G_d(p)/L$ and the analytic orbit-reference
reversal threshold, with pointwise 95\% block-bootstrap intervals on the losses.
The $d/L=0$ and orbit endpoints are analytic.
(b) $I_3$ with trajectory standard errors. The two vertical lines mark
the limits of $p_c=0.159(3)$, including finite-size drift.
(c) Peak of $(\overline S_{C_L}-\overline S_R)/L$ versus $1/L$ across
measurement rates, with $\overline S_R$ the mean for the complete rainbow
$\ket{\Phi_n^0}$. Error bars show
standard errors from peaks recomputed in each paired bootstrap resample.
Lines guide the eye.}
\label{fig:SM_pilots}
\end{figure}

For fixed $0.065<p<1/4$ and sufficiently large allowed $L$, the first-readout
reversal density $(3/8)[1-(1-p)^2(1+p^2/5)]-3/64$ is positive and the
orbit-reference loss density vanishes. The original rainbow--Clifford
comparison [Fig.~\ref{fig:SM_pilots}(c)] has 2000 paired trajectories per
cell at $L=64,128,256,512,1024$, $p=0.08,0.12,0.16,0.20,0.24$;
in each of 2000 paired bootstrap resamples, the peak is recomputed
on $t=0,0.5,\ldots,4$.

\subsection{Haar dynamics and nonflat spectra}
\label{sec:SM_nonclifford}
Both state-vector controls use $L=16$, an $8|8$ cut, $p=0.20$ and
$t=0,0.5,\ldots,8$. Alternating open-chain Haar $U(4)$ gates
\cite{Mezzadri2007How}, starting on odd bonds, precede conditional-Born
Bernoulli-$Z$ readout after each half-layer. Preparation uses within-half
Haar gates without readout. Within each replica, all conditions share
gates, masks and the uniform random numbers used to sample outcomes,
with each input using its own conditional Born probabilities.

\paragraph{Flat-spectrum results.}
The flat-spectrum sample has 100 independent blocks with 20 replicas per block.
The nested preparations $\ket{\Phi_6^d}$, $d=0,2,4,8$, share the
independently encoded reference $\ket{\Phi_5^8}$, so $D_d(0)=-1$.
In each of ten thousand paired block bootstrap resamples, we recompute
$\Delta_d=\max_{t=0,0.5,\ldots,8}D_d(t)$. These resamples give two-sided 95\%
percentile intervals for peaks, adjacent contrasts, losses and bare $\chi$.
Centered suprema over $t=0.5,1,\ldots,8$ give nominal per-depth 95\%
bands with half-widths of approximately $0.034$, $0.023$, $0.021$, and
$0.021$ ebits in order of increasing depth.
\begin{center}
\begin{tabular}{rrrr}
Preparation & $\Delta_d$ [95\% interval] & First-readout loss & Bare $\chi$\\\hline
$\ket{\Phi_6^0}$ & $0.630\;[0.607,0.654]$ & 2.116 & 12\\
$\ket{\Phi_6^2}$ & $0.239\;[0.226,0.252]$ & 1.744 & 10.96\\
$\ket{\Phi_6^4}$ & $0.058\;[0.049,0.069]$ & 1.510 & 9.10\\
$\ket{\Phi_6^8}$ & $0.003\;[-0.006,0.022]$ & 1.179 & 5.45\\
$\ket{\Phi_5^8}$ & --- & 0.722 & 3.06\\\hline
\end{tabular}
\end{center}
The first resolved crossing occurs at $t=0.5$ for $d=0$ and at $t=1$ for
$d=2,4$. No crossing is resolved for $d=8$ on this grid.
For successive depth pairs $(0,2)$, $(2,4)$, and $(4,8)$, the peak
decreases by $0.391\;[0.368,0.415]$, $0.182\;[0.167,0.192]$, and
$0.055\;[0.035,0.068]$ ebits, respectively (95\% intervals). Bare $\chi$ sums one-site Born-averaged
losses. With $G_d=6-\overline S_d(1/2)$, $G_{\rm ref}=5-\overline S_{\rm ref}(1/2)$,
$D_d(1/2)=G_d-G_{\rm ref}-1$; $D_4(1/2)\simeq-0.213$ ebit before the crossing.

\begin{figure}[!htbp]
\centering\includegraphics{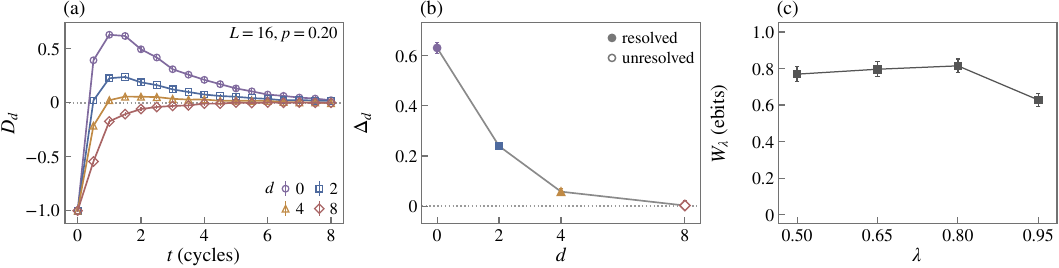}
\caption{Haar controls at $L=16$, $p=0.20$.
(a,b) Flat spectra (100 blocks, 20 replicas per block): mean differences $D_d(t)$
with nominal per-depth 95\% simultaneous bands (a), and peaks $\Delta_d$
with 95\% bootstrap intervals (b). Filled (open) markers in (b) indicate
crossings resolved (unresolved) by the simultaneous bands in (a).
(c) Early ordering contrast $W_\lambda$ between $d=0$ and $d=8$
[Eq.~\eqref{eq:nonflat-targets}], from 64 blocks with two replicas per block.
The spectrum is flat at $\lambda=0.50$. Error bars are block standard
errors; lines guide the eye. Table~\ref{tab:nonflat-four} gives simultaneous
lower bounds.}
\label{fig:SM_nonclifford}
\end{figure}

\paragraph{Four-spectrum results.}\label{sec:SM_nonflat_confirmation}
For $\lambda=0.50,0.65,0.80,0.95$, seeds have six (high) or five
(reference) outer pairs
$\sqrt\lambda\ket{00}+\sqrt{1-\lambda}\ket{11}$ and $\ket{0}$
elsewhere, with entropies $6h_2(\lambda)$ and $5h_2(\lambda)$,
$h_2(x)=-x\log_2x-(1-x)\log_2(1-x)$.
Nested high-state depths $d=0,2,4,8$ preserve the full spectrum;
the independent reference has depth 8. Encoders are independent on each
half and reused across $\lambda$. All 20 conditions share each replica's
monitoring. The 64 independent blocks have two replicas each and are separate
from the flat-spectrum study.
At fixed $\lambda$, we test $G_0-G_8$, $D_0(1)$ and
\begin{equation}
 W_\lambda=\frac14\sum_{t=0.5,1,1.5,2}[D_0(t)-D_8(t)].
 \label{eq:nonflat-targets}
\end{equation}
Reference cancellation makes the first-readout loss reduction and
$W_\lambda$ depend only on high-state entropies. Each block supplies a
12-component row. As initial high-state entropy falls from $6$ to $1.72$
ebits, $W_\lambda$ changes from $0.769$ to $0.628$ ebit
[Fig.~\ref{fig:SM_nonclifford}(c)]. All twelve lower bounds in
Table~\ref{tab:nonflat-four} are positive: depth 8 reduces first-readout
loss and early ordering difference, and the unencoded preparation reverses
mean order by $t=1$ at every spectrum.

\begin{table}[!htbp]
\centering
\caption{Four-spectrum means (block standard errors) and one-sided simultaneous lower bounds at nominal 95\% confidence, in ebits. Bounds are rounded downward; $c_{\rm boot}\simeq2.411$.}
\label{tab:nonflat-four}
\begin{tabular}{crrrrrr}\hline
 & \multicolumn{2}{c}{$G_0-G_8$} & \multicolumn{2}{c}{$W_\lambda$} & \multicolumn{2}{c}{$D_0(1)$}\\
$\lambda$ & Mean (SE) & Lower & Mean (SE) & Lower & Mean (SE) & Lower\\\hline
0.50 & 0.976 (0.065) & 0.820 & 0.769 (0.042) & 0.667 & 0.590 (0.054) & 0.458\\
0.65 & 0.977 (0.060) & 0.832 & 0.796 (0.040) & 0.699 & 0.624 (0.054) & 0.493\\
0.80 & 0.889 (0.051) & 0.766 & 0.815 (0.038) & 0.723 & 0.664 (0.052) & 0.538\\
0.95 & 0.413 (0.030) & 0.342 & 0.628 (0.034) & 0.545 & 0.593 (0.045) & 0.484\\\hline
\end{tabular}
\end{table}

\paragraph{Simultaneous inference.}\label{sec:SM_nonflat_statistics}
Block standard deviations $s_j$ use divisor 63, with SE $s_j/8$.
Each of 49,999 paired resamples retains all twelve coordinates of 64
whole rows and recomputes means and standard deviations. For support
endpoint $a_j$, define
\begin{equation}
 T_j^*=8(\bar X_j^*-\bar X_j)/s_j^*,\quad
 M^*=\max_jT_j^*,\quad c_{\rm boot}=\max\{M^*_{(47500)},0\},\quad
 L_j^{\rm boot}=\max\{a_j,\bar X_j-c_{\rm boot}s_j/8\}.
 \label{eq:nonflat-bootstrap-rule}
\end{equation}
Here $M^*_{(47500)}$ is the 47,500th smallest bootstrap maximum
without interpolation. All marginal standard deviations were positive. For iid bounded rows of
fixed dimension and positive variances, coverage is nominal 95\%
asymptotically and simultaneous \cite[Sec.~4.2]{Romano2003Stepdown}.
Supports for $G_0-G_8,W_\lambda,D_0(1)$ are respectively
$[-6,6]$, $[-7.5,7.5]$, $[-8,7]$. For the first two, reference cancellation
leaves high-state differences: rank at most $2^6$ after the first intra-half
readout and entropy at most eight thereafter give $(6+8+8+8)/4=7.5$.
At $t=1$, cross-cut operator Schmidt rank at most four bounds reference
rank by $2^7$ and high-state entropy by eight, on every branch and spectrum.

\label{page:sm-end}
\end{document}